\documentclass[12pt]{article}
\usepackage{epsf}

\newcommand{\bmat}{\left(\begin{array}}
\newcommand{\emat}{\end{array}\right)}

\def\yzero{\smash{\hbox{$y\kern-4pt\raise1pt\hbox{${}^\circ$}$}}}

\def\a{\alpha}
\def\b{\beta}
\def\g{\gamma}
\def\d{\delta}
\def\beq{\begin{equation}}
\def\eeq{\end{equation}}
\def\beqa{\begin{eqnarray}}
\def\eeqa{\end{eqnarray}}
\def\Om{\Omega}
\def\om{\omega}
\def\th{\theta}
\def\vt{\vartheta}

\def\-{\hphantom{-}}
\def\ov{\overline}
\def\s2{\frac{1}{\sqrt2}}

\def\oh{\frac{1}{2}}
\def\beq{\begin{equation}}
\def\eeq{\end{equation}}
\def\beqa{\begin{eqnarray}}
\def\eeqa{\end{eqnarray}}

\def\IF{\relax{\rm I\kern-.18em F}}
\def\II{\relax{\rm I\kern-.18em I}}
\def\IP{\relax{\rm I\kern-.18em P}}
\def\IC{\relax\hbox{\kern.25em$\inbar\kern-.3em{\rm C}$}}
\def\IR{\relax{\rm I\kern-.18em R}}

\def\cn{{\cal N}}
\def\cam{{\cal M}}

\def\Dsl{\,\raise.15ex\hbox{/}\mkern-13.5mu D} 
\def\IZ{Z\kern-.4em  Z}

\def\A{{\bf A}}
\def\B{{\bf B}}
\def\ent{{\bf Z}}
\def\C{{\bf C}}
\def\ti{\times}
\def\OR{\Omega {\cal R}}
\def\R{{\cal R}}
\def\ca{{\cal A}}
\def\lam{\lambda}
\def\raw{\rightarrow}

\def\G{\Gamma}
\def\ep{\epsilon}
\def\arr{\arrowvert}

\catcode`\@=11
\newdimen\@rotdimen
\newbox\@rotbox

\def\@vspec#1{\special{ps:#1}}
\def\@rotstart#1{\@vspec{gsave currentpoint currentpoint translate
   #1 neg exch neg exch translate}}
\def\@rotfinish{\@vspec{currentpoint grestore moveto}}
\def\@rotr#1{\@rotdimen=\ht#1\advance\@rotdimen by\dp#1%
   \hbox to\@rotdimen{\hskip\ht#1\vbox to\wd#1{\@rotstart{90 rotate}%
   \box#1\vss}\hss}\@rotfinish}
\def\@rotl#1{\@rotdimen=\ht#1\advance\@rotdimen by\dp#1%
   \hbox to\@rotdimen{\vbox to\wd#1{\vskip\wd#1\@rotstart{270 rotate}%
   \box#1\vss}\hss}\@rotfinish}%
\def\@rotu#1{\@rotdimen=\ht#1\advance\@rotdimen by\dp#1%
   \hbox to\wd#1{\hskip\wd#1\vbox to\@rotdimen{\vskip\@rotdimen
   \@rotstart{-1 dup scale}\box#1\vss}\hss}\@rotfinish}%
\def\@rotf#1{\hbox to\wd#1{\hskip\wd#1\@rotstart{-1 1 scale}%
   \box#1\hss}\@rotfinish}%
\def\rotate{\@ifnextchar[{\@rotate}{\@rotate[l]}}
\def\@rotate[#1]#2{\setbox\@rotbox=\hbox{#2}\@nameuse{@rot#1}\@rotbox}

\catcode`\@=12

\topmargin
-1.5cm
\textwidth
15.5cm
\textheight
23.5cm
\oddsidemargin
0.7cm
\evensidemargin
1.2cm

\begin{document}

\makeatletter
\@addtoreset{equation}{section}
\makeatother
\renewcommand{\theequation}{\thesection.\arabic{equation}}
\pagestyle{empty}
\rightline{FTUAM-02/12; IFT-UAM/CSIC-02-13}
\rightline{\tt hep-th/0205074}
\vspace{0.5cm}
\begin{center}
\LARGE{ Standard Model at Intersecting D5-branes : Lowering the 
String Scale \\[10mm]}
\large{
D. Cremades, L.~E.~Ib\'a\~nez and  F. Marchesano  
\\[2mm]}
\small{
 Departamento de F\'{\i}sica Te\'orica C-XI
and Instituto de F\'{\i}sica Te\'orica  C-XVI,\\[-0.3em]
Universidad Aut\'onoma de Madrid,
Cantoblanco, 28049 Madrid, Spain.
\\[9mm]}
\small{\bf Abstract} \\[7mm]
\end{center}

\begin{center}
\begin{minipage}[h]{14.0cm}
Recently a class of Type IIA orientifold models was constructed 
yielding just the fermions of the SM at the intersections of D6-branes
wrapping a 6-torus. We generalize that construction to the case of 
Type IIB compactified on an orientifold of 
${\bf T^4 \times (C/Z_N)}$ with D5-branes intersecting at 
angles on ${\bf T^4}$. 
We construct explicit models in which the
massless fermion spectrum is just the one of a three-generation
Standard Model. 
 One of the motivations for these 
new constructions is that  
 in this case there are  2 dimensions which are
transverse to the SM D5-brane configuration. By making those
two dimensions large enough one can have a low string scale
$M_s$ of order 1-10 TeV and still have a large $M_{Planck}$  
in agreement with observations. 
From this point of view,
these are the first explicit D-brane string constructions 
where one can achieve having just the fermionic spectrum and
gauge group of the SM embedded in a Low String Scale
scenario. The cancellation
of $U(1)$ anomalies turns out to be quite analogous to the toroidal
D6-brane case and the proton is automatically stable due to the gauging
of baryon number. Unlike the D6-brane case, the present class of models
has $\cn = 0$ SUSY both in the bulk and on the branes and hence the spectrum
is simpler.

\end{minipage}
\end{center}
\newpage
\setcounter{page}{1}
\pagestyle{plain}
\renewcommand{\thefootnote}{\arabic{footnote}}
\setcounter{footnote}{0}


\section{Introduction}

The brane-world idea has become popular in the last couple of years.
 In this scheme 
it is assumed that the standard model (SM) fields and
interactions are confined to a $(p+1)$ dimensional submanifold
of a larger D-dimensional ($D>(p+1)$) manifold in which 
gravitational fields propagate. Dp-branes in string theory
provide a natural setting in which this scenario arises,
since gauge interactions are confined to the world-volume
of branes. However, in the brane-world scenario studies
a crucial property of the SM is often ignored, the fact 
that its spectrum is chiral. Dp-branes on a smooth 
space have non-chiral extended SUSY on their worldvolume.

In order to obtain explicit D-brane realizations of the SM 
it is thus necessary to do something to obtain chiral 
theories on their worldvolume. One of the simplest options 
to obtain chirality is locating stacks of branes on
top of some, e.g., orbifold singularity on transverse 
space. For example, three generation models may be obtained
by locating sets of D3-branes on top of a $\ent_3$ singularity 
\cite{bbarmod,aiqu} (see also \cite{cuw,bailind4,bjl,auquivers,
kephart,hxy}).
Local tadpole cancellation in general requires the immersion of 
those D3-branes on some D7-branes. These are simple theories
with phenomenological interest. However the  spectrum in general
goes beyond the minimal content of the SM (or the MSSM), since
extra doublet fermions appear in the spectrum due to the 
structure of $U(1)$ anomaly cancellation
\footnote{For attempts to obtain models of D3-branes 
on a $\ent_N$ singularity without extra fermionic doublets see
ref.\cite{gerardo}.}.

Another option in order to get $D=4$ chirality is to consider
intersecting D-branes \cite{bdl,arfaei}
(for somewhat related work see also \cite{bachas,inter1,fluxes}).  
Recently, a class of intersecting D-brane configurations yielding just the
fermionic spectrum of the SM was for the first time constructed
\cite{imr}.
They are obtained from four stacks  of D6-branes wrapping 
an orientifolded 6-torus and intersecting at angles
\cite{bgkl,afiru,afiru2,bkl} (see also
\cite{pseudogermans,bklo,csu,pheno,honecker,kataoka,cim1,cim2,koko}
for further developments).
In the bulk there is $\cn = 4$ supersymmetry but the spectrum
is chiral at the brane intersections. The models are 
in general non-supersymmetric but for certain choices 
of the compact radii one can preserve some $\cn = 1$ SUSY
at each intersection \cite{csu,cim1,cim2}.  
One of the nice features of the simplest such
constructions is that the existence of  three 
quark-lepton generations is related 
(via $U(1)$ anomaly cancellation)
to the presence  of three
colours in QCD \cite{imr}. Another interesting feature is that one
may expect the appearance of some exponential suppression
in certain  Yukawa couplings, providing a means to understand
the hierarchical structure of fermion masses \cite{afiru2}.
The SM Higgs mechanism has an interpretation as a brane recombination
process in which the branes supporting the SM gauge group 
are recombined into a single brane related to electromagnetism
\cite{afiru2,cim2}.

One point which remains to be understood in those 
brane configurations is the hierarchy between the Planck scale 
and the weak scale. The models are non-supersymmetric and
in order to avoid the standard gauge hierarchy problem
of the Higgs scalars a natural  option is to assume the 
string scale $M_s$ not much above the electroweak
scale, i.e., $M_s \sim 1-10$ TeV. Then a possibility 
for understanding the observed smallness of gravitational 
interactions would be to have some compact dimensions
(transverse to the brane)  
very large, as suggested in \cite{TeV,aadd}. However in the 
case of these intersecting D6-brane models one can see 
there is no compact direction transverse to all SM
stacks of branes \cite{bgkl}. Thus one has to look for other 
possible sources of suppression for gravitational 
interactions like e.g., that in ref.\cite{rs}.

A natural alternative is
to  consider  instead of D6-branes lower dimensional
ones,  intersecting D5- and D4-branes. This would be interesting since,
as pointed in ref.\cite{afiru},  
then there are more transverse dimensions to the branes which
can be made large, allowing for a low string scale 
$M_s<<M_p$.   
In the present paper we extend the work of ref.\cite{imr}
to the case 
of intersecting D5- and D4-branes  wrapping cycles
on ${\bf T^2\times T^2\times T^2/\ent_N}$
and ${\bf T^2\times (T^4/\ent_N)}$ respectively
\footnote{ Unlike the case of D6-branes, D5- or D4-branes wrapping 
cycles on ${\bf T^2\times T^2\times T^2}$ do not lead to D=4 chiral theories.
This is why in order to achieve chirality an additional 
$\ent_N$ twist in transverse dimensions  
is performed in the constructions of the present paper.}.
In the case of D5-branes, these are localized on a
fixed point of the  
orbifold ${\bf T^2/\ent_N}$ and wrap 2-cycles on ${\bf T^2\times T^2}$.
This class of constructions was already introduced in \cite{afiru},
but in order to obtain just the spectrum of the 
SM we will be now considering orientifolds of such 
constructions. 

We will be able to obtain intersecting D5-brane models with
the fermionic spectrum and gauge group of the SM.
The $U(1)$ anomaly structure is identical to  
that of the D6-brane  models of ref.\cite{imr}. There are however 
many differences between both classes of models. The present class
of models have $\cn = 0$ SUSY both in the bulk and on the branes
and none of the quarks, leptons or gauge bosons have any
SUSY partner. Thus the massless spectrum is closer to that
of the non-SUSY SM spectrum. The only light particles beyond the
SM spectrum will be some extra scalars (often coloured) and
a minimal  Higgs system analogous to that of the MSSM.

The structure of the paper is as follows. In the next section 
we describe the construction of intersecting D5-branes 
wrapping cycles on ${\bf T^2\times T^2\times C/\ent_N}$.
We derive the RR-tadpole cancellation conditions for the
orientifold case and obtain the lightest spectrum. 
The cancellation of mixed $U(1)$ anomalies through a
generalized Green-Schwarz mechanism is presented
in some detail. In Section 3 we present the general strategy to obtain 
intersecting  D5-brane configurations with the spectrum of the SM. 
We present a particular example in some detail and leave further
examples for Appendix II. Examples of left-right symmetric models 
are  provided as well. We also show how to construct a left-right 
symmetric model free of open string tachyons for any odd value 
$N$ of the twist $\ent_N$. Some general physical issues like 
the form of the lightest  spectrum beyond the SM and
the lowering of the string scale are discussed in Section 4.
We leave some general comments and conclusions for Section 5.

In Appendix I we analyze the case of
intersecting D4-branes wrapping one-cycles 
on ${\bf T^2\times T^4/\ent_N}$. These configurations turn out to be less
flexible for model building purposes. In particular
there is no obvious  way to  obtain just the SM fermion spectrum 
at the intersections. We nevertheless provide some
semi-realistic example based on intersecting D4-branes 
in that appendix.

\section{Intersecting D5-branes on ${\bf T^4\times C/\ent_N}$ orientifolds}

Let us describe the general intersecting branes setup 
that we will be considering. As was explained in \cite{afiru},
chiral compactifications may naturally arise from considering 
configurations of D($3+n$)-branes filling four-dimensional
Minkowski space, wrapping $n$-cycles of a 
$2n$-dimensional compact manifold $\A_{\bf 2n}$ and sitting at a point 
in a transverse ($6-2n$)-dimensional manifold $\B_{\bf 6-2n}$. In order
to obtain a chiral spectrum from the open string sector, 
the cycles the branes wrap should have nontrivial intersection 
numbers, while the point they sit in $\B_{\bf 6-2n}$ must be singular.
Lowering the string scale in a natural way implies, in turn,
having $n < 3$, so that we can consider a nontrivial transverse 
space $\B_{\bf 6-2n}$ whose global properties (as its volume) do not 
directly affect our open string sector (where our chiral gauge theory arises),
but only the closed string sector.

The special case $n = 0$, that corresponds to D3-branes
on top of a singularity, was already considered in \cite{aiqu}
(and more recently in \cite{gerardo}), yielding semi-realistic
gauge theories in $D=4$. The cases $n = 1,2,3$ were then considered
in \cite{afiru,afiru2} in a simple setup where ${\bf A_{2n} = T^{2n}}$ 
and the branes sit in an orbifold singularity $\C^{3-n}/\ent_N$.
However, as was explicitly shown in \cite{imr}, considering
orientifold models may greatly simplify our chiral spectrum, being 
possible to attain configurations where the matter content
just reduces to the SM fermion content. These models were 
constructed in a particular setup of the case $n = 3$, where
D$6$-branes wrap 3-cycles of ${\bf A_{2n} = T^2 \ti T^2 \ti T^2}$,
as previously considered in \cite{bgkl,bkl}.

This fact naturally lead us to consider orientifold models 
of branes at angles. In particular, we will consider the 
orientifold version of the compactifications already constructed
in \cite{afiru}, focusing on the cases $n = 1,2$ that allow us
to obtain  low string scale scenarios \cite{TeV,aadd}. Some related 
constructions of branes at angles have been considered in 
\cite{honecker,kataoka}. Notice, however, that the class of models
constructed in the present paper are more general, in the sense
that, following the {\it Bottom-Up} approach described in \cite{aiqu},
we will only bother about the local physics arising from the
singular point in $\B_{\bf 6-2n}$ where the D-brane content lies. 
The specific models presented as explicit examples of 
such constructions are also new, as well as their associated
phenomenology.

\subsection{Construction}

Let us consider some specific D-brane setting where the
above scenario can be naturally realized. As previously
stated, this will imply considering configurations of 
D5(D4)-branes wrapping 2(1)-cycles of a 4(2)-dimensional
compact manifold $\A$ 
which is in turn embedded in some 6-dimensional manifold $\cam$
as the `tip' of an orientifold singularity.
We will consider in what follows the D5-brane
case, leaving the discussion of D4-brane constructions  
for an appendix. Following \cite{afiru}, we will choose 
a fairly simple realization of this setup, given by
\beq
{\rm Type \ IIB \ on \ } M_4 \ti 
\frac{\ T^{4} \ti \C/\ent_N}{\{1 + \OR\}},
\label{singuori}
\eeq
where $\R$ stands for an involution associated with the parity
reversal operation $\Om$. In our case, 
$\R = \R_{(5)}\R_{(7)}\R_{(8)}\R_{(9)} (-1)^{F_L}$,
$\R_{(i)}$ standing for a reflection in the $i^{th}$ coordinate
and ${F_L}$ being the left fermion number.
More specifically, 
if we describe our internal coordinates by complex variables
$Z_i = X_{2i+2} + iX_{2i+3}$, then  
${\cal R}$ is given by the geometrical action
\beq
\begin{array}{rlc}
{\cal R}_g : & Z_i \longmapsto \bar Z_i, & i = 1,2 \\  
& Z_3 \longmapsto - Z_3.
\end{array}
\label{R5}
\eeq

\begin{figure}
\centering
\epsfxsize=4.4in
\hspace*{0in}\vspace*{.2in}
\epsffile{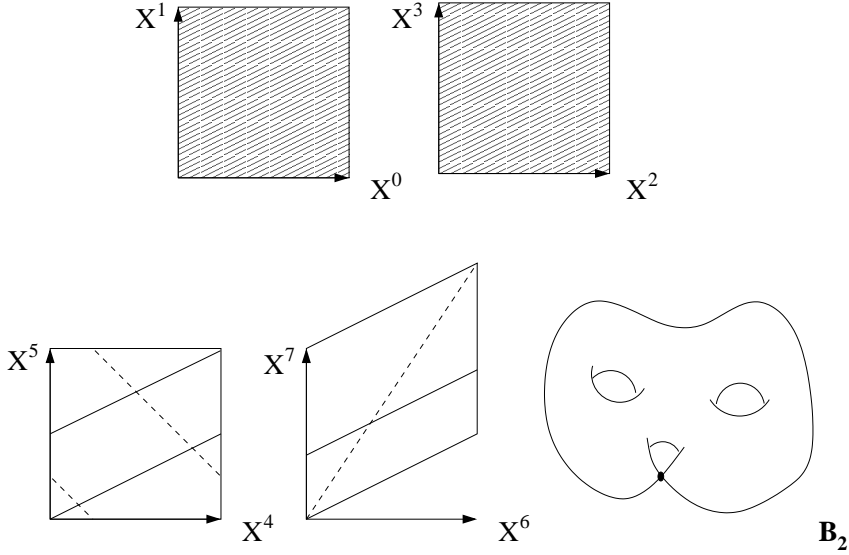}
\caption{\small{Intersecting brane world setup.
We consider configurations of D5-branes filling four-dimensional
Minkowski spacetime, wrapping factorizable 2-cycles of ${\bf T^2 \ti T^2}$
and sitting at a singular point of some compact two-dimensional space
$\B_2$. In the figure, two such branes are depicted, with
wrapping numbers $(1,2)(1,\frac 32)$ (solid line) and $(1,-1)(1,\oh)$
(dashed line). The fractional wrapping numbers arise from a
tilted complex structure: $b^{(1)} = 0$, $b^{(2)} = \oh$.
}
\label{world}}
\end{figure}

Such orientifold singularity will induce, as usual, a non-vanishing 
Klein-bottle amplitude, signaling the presence of an O5-plane 
in our configuration. In order to cancel the associated 
RR tadpoles, we will introduce $K$ stacks of $N_a$ D5-branes 
filling $M_4$ and wrapping 2-cycles $[\Pi_a] \in H_2 (T^4,\ent)$
($a = 1,\dots,K$), while sitting at the origin of $\C/\ent_N$,
$N$ being an odd integer
\footnote{In the compact case of interest the D5-branes will be sitting 
at a generic $\ent_N$ singularity in the 
third complex compact dimension.} . 
Furthermore, we will consider a particularly 
simple subclass of configurations where ${\bf T^4}$ is a factorizable
torus ${\bf T^2 \ti T^2}$, and the 2-cycles the branes wrap can be decomposed
as a product of two 1-cycles $[(n_a^1, m_a^1)] \otimes [(n_a^2, m_a^2)]$, 
each wrapping a different ${\bf T^2}$ (see figure \ref{world} for an example). 

The $\ent_N$ orbifold twist on the third complex dimension is 
generated by a geometric action $\om$, encoded 
in a twist vector of the form $v_\om = \frac 1N (0,0,-2,0)$
for modular invariance requirements and for
the variety to be spin. This same $\ent_N$ action may be 
embedded in the $U(N_a)$ degrees of freedom arising from the
$a^{th}$ stack of D5-branes, through a unitary matrix of the 
form
\beq
\g_{\om,a} = {\it diag} \left( {\bf 1}_{N_a^0},\ \a {\bf 1}_{N_a^1}, 
\ldots,\ \a^{N-1} {\bf 1}_{N_a^{N-1}} \right),
\label{Chan}
\eeq
with $\sum_{i=0}^{N-1} N_a^i = N_a$, and where we have defined 
$\a \equiv {\rm exp} (2\pi i/N)$.

This same class of configurations can be analyzed in a T-dual
picture, in terms of Type IIB D7-branes with non-trivial 
wrapping numbers and fluxes in the first two compact 
complex dimensions, while again localized 
in the orientifold singularity.
Furthermore, when embedding such singularity in a simple toroidal 
orbifold as ${\bf T^2/\ent_N}$, any configuration can be easily 
related to Type I compactified on ${\bf T^2 \ti T^2 \ti T^2/\ent_N}$, 
with some $F$ and $B$-fluxes in the first two tori.
As discussed in \cite{bkl} (see also \cite{bflux}), in such compactifications
only discrete values of the $b$-field are allowed by the 
presence of $\Om$, namely $b = 0, \oh$. In our T-dual
picture of branes at angles, this can be seen by 
noticing that the geometric action of $\R$ 
restricts the generic toroidal complex structures of
${\bf T^2 \ti T^2}$ to those that are invariant under complex conjugation.
This allows us to consider either rectangular ($b = 0$) 
or special tilted ($b = \oh$) lattices when defining our ${\bf T^2}$
(see figure \ref{world}). In order to describe
configurations with non-vanishing $b$, is convenient to define 
effective wrapping numbers
\beq
(n_a^i,m_a^i)_{eff} := (n_a^i,m_a^i) + b^{(i)}(0,n_a^i), 
\label{frac}
\eeq
where $b^{(i)}$ stands for the value of $b$ on the $i^{th}$ torus 
${\bf T^2}$. This simple redefinition of the wrapping numbers allows
us to simply describe the action of $\OR$ in the open string sector. 
Indeed, in order to have a consistent compactification we should
always consider either D5-branes invariant under $\OR$ or
pairs of branes related by its action. If a D$5_a$-brane is 
described by 
\beqa
& (n_a^1,m_a^1) \otimes (n_a^2,m_a^2) \nonumber \\
& \g_{\om,a} = {\it diag} \left( {\bf 1}_{N_a^0},\ \a {\bf 1}_{N_a^1}, 
\ldots,\ \a^{N-1} {\bf 1}_{N_a^{N-1}} \right),
\label{brana_a}
\eeqa
then the sector $\OR$D$5_a$ or D$5_{a^*}$ will be given by
\beqa
& (n_a^1,-m_a^1) \otimes (n_a^2,-m_a^2) \nonumber \\
& \g_{\om,a^*} = {\it diag} \left( {\bf 1}_{N_a^0},\ \a^{N-1} 
{\bf 1}_{N_a^{1}}, \ldots,\ \a {\bf 1}_{N_a^{N-1}} \right),
\label{brana_a*}
\eeqa
where we consider the effective wrapping numbers defined in (\ref{frac}).
Both branes $a$ and its mirror partner $a^*$ should be included in 
a consistent configuration. 

Let us now analyze the low energy spectrum arising from such generic 
class of configurations. First let us consider the closed string sector,
which can be computed using standard orbifold techniques.
Such techniques have been recently used in a sistematic 
study of non-supersymmetric Type II and Heterotic toroidal orbifolds
in \cite{anamaria}. In particular, the closed Type IIB
spectrum has been explicitely computed for the toroidal embedding 
of the $N = 3$ orbifold singularity, so we refer the reader to the 
appendix of \cite{anamaria} for further details. 

Let us first notice that, since we are only concerned with the 
physics arising at the orientifold singularity, 
it is pointless to compute the {\it untwisted} 
sector of the theory, which concerns the whole compactification. 
However, when embedding this $\ent_N$ singularity in a compact 
six-dimensional manifold $\cam$, this sector should give rise to 
four dimensional gravitation plus some other extra massless 
particles. At this level we will only  state that, since
the twist $v_\om$ is explicitly non-supersymmetric, 
the spectrum in the bulk will necessarily present $\cn = 0$,
thus yielding a more economical spectrum that
the one obtained by plain dimensional reduction on a torus.

On the contrary, the closed string {\it twisted} sector
of the theory will play a relevant role with respect to
the local physics on the singularity. As expected, 
Type IIB theory on such singularity will give rise to 
RR twisted $p$-forms of even $p$:
\beq
\begin{array}{cccc}
A_0^{(k)}, & A_2^{(k)}, & A_4^{(k)}, & A_6^{(k)}
\end{array}
\label{pformas}
\eeq
where $k = 1, \dots, N-1$ denotes the $k^{th}$-twisted sector of the theory.
$A_p^{(k)}$ and $A_{6-p}^{(k)}$ field strenghts are related
by Hodge duality in $D = 8$, while the orientifold action 
identifies $k$ and $N - k$ sectors. 
In addition, each NSNS $k^{th}$-twisted sector will lead to
a closed string tachyon of $\a'$(mass)$^2 = - \frac {4 k}{N}$
(actually, due to the $\OR$ modding, only $\oh (N-1)$ {\em real} 
such tachyons do actually appear).
The physical interpretation  of analogous  type of closed string tachyons
have been recently discussed in ref.\cite{tachyons}
for the case of non-compact orbifolds. A similar discussion in the
case of compact orbifolds and orientifolds 
is still lacking and goes beyond
the scope of the present paper.
Note in particular that in the ${\bf C/Z_N}$  case 
considered in \cite{tachyons} the tachyons are complex 
and their vev signal the smoothing of the singularity.
In the present {\em orientifold } case the tachyons are real fields
and the analysis should be different.

Let us now focus on the open string sector of the theory.
Whenever a D$5_a$-brane is not invariant under the orientifold 
action $\OR$, the massless spectrum arising from the 
D$5_a$D$5_a$ sector is identical to the one computed 
for the orbifold case in \cite{afiru}, since this sector will be
mapped to the D$5_{a^*}$D$5_{a^*}$ sector, and there will not be
any associated $\Om$ projection. 
This spectrum can be easily described in bosonic language.
Indeed, to each open string excitation we can associate 
a four-dimensional vector $r \in (\ent + \nu)^4$, where
$\nu$ distinguishes between the Ramond ($\nu = \oh$) 
and Neveu-Schwarz ($\nu = 0$) sectors of the theory.
The GSO projection is imposed by requiring $\sum_i r^i =$ odd,
and the massless states are those that satisfy $\sum_i (r^i)^2 = 1$.
Namely, the massless states surviving the 
GSO projection in both R and NS sectors are
{\small \beqa
\begin{array}{cccc}
{\rm \bf NS\ State} \quad & \quad {\bf \ent_N \ phase}  \quad 
& \quad {\rm \bf R \ State} \quad & \quad {\bf \ent_N \ phase} \\
(\pm1,0,0,0) & 1 & \pm\oh(-,+,+,+) & e^{\mp2\pi i \frac{1}{N}}\\
(0,\pm1,0,0) & 1 & \pm\oh(+,-,+,+) & e^{\mp2\pi i \frac{1}{N}}\\
(0,0,\pm1,0) & e^{\mp4\pi i \frac{1}{N}} & \pm\oh(+,+,-,+) 
& e^{\pm2\pi i \frac{1}{N}}\\
(0,0,0,\pm1) & 1 & \pm\oh(+,+,+,-) &  e^{\mp2\pi i \frac{1}{N}}
\label{sector5aa}
\end{array}
\eeqa}
where its behaviour under the $\ent_N$ orbifold twist has also 
been indicated. As usual, the open string spectrum is computed
by keeping states invariant under the combined geometric plus 
Chan-Paton (CP) $\ent_N$ action \cite{dg}, so after this projection
we are led to a spectrum of the form
{\small\beq
\begin{array}{rl}
{\rm\bf Gauge\; Bosons} & \prod_a \prod_{i=1}^N U(N_a^i)
\nonumber\\
{\rm\bf Complex \; Scalars} & \sum_a \sum_{i=1}^N
[\; (N_a^{i},{\ov N}_a^{i-2}) + 2\times  {\bf Adj}_a^i\; ] \nonumber\\
{\rm\bf Left\; Fermions} 
& \sum_a \sum_{i=1}^N 2 \ti (N_a^{i},{\ov N}_a^{i-1}) \nonumber\\
{\rm\bf Right\; Fermions} 
& \sum_a \sum_{i=1}^N 2 \ti (N_a^{i},{\ov N}_a^{i-1})
\label{espectro5aa}
\end{array}
\eeq}
where the index $i$ is defined $mod$ $N$.
Notice that this spectrum is explicitly 
non-chiral and also non-supersymmetric,
since all the gauginos have been projected out. Notice,
as well, that when considering D5-branes invariant under 
the $\OR$ action, $SO(N)$ and $USp(N)$ gauge groups will
also arise. Since we are not interested in constructing
configurations with these groups, we will not consider
this option any longer.

Both the chiral matter and the tachyonic content of our 
configuration will arise from the sectors $D5_aD5_b$,
$D5_aD5_{b^*}$ and $D5_aD5_{a^*}$. Let us compute this spectrum
explicitly for the $D5_aD5_b$ sector. Just as in the
$D5_aD5_a$ case, this sector is not constrained by the 
$\OR$ projection, so its associated spectrum is computed
in the same way as in an orbifold compactification.
In order to properly describe it, we can introduce
the twist vector $v_\vt = (\vt_{ab}^1, \vt_{ab}^2, 0, 0)$,
$\pi \vt_{ab}^i$ being the angle that both branes form
on the $i^{th}$ torus \cite{afiru}.
The states living in the $ab$ intersection are then
characterized by four-dimensional vectors of the form
$r + v_\vt$, where $r$ stands for the set of discrete
vectors introduced above.  The mass formula is also modified
to \cite{arfaei,afiru}
\beq
\a' M_{ab}^2 = {Y^2 \over 4\pi\a^\prime} + N_{bos}(\vt) 
+ {(r + v_\vt)^2 \over 2} -\oh + E_{ab},
\label{mass2}
\eeq
where $Y$ stands for any transversal separation between both branes,
$N_{bos}(\vt)$ is the bosonic oscillator contribution and $E_{ab}$ is
the vacuum energy:
\beq
E_{ab} = \sum_{i=1}^3 \oh |\vt^i| (1 - |\vt^i|)
\label{vacio}
\eeq

The massless and tachyonic states will now be
{\small\beq
\begin{array}{cccc}
{\rm\bf Sector} & {\rm\bf State} & {\rm\bf \ent_N \ phase} & {\rm\bf \a' Mass^2}
\vspace{2mm}\\
{\rm NS} & (-1+\vartheta^1,\vt^2,0,0) & 1 & -\oh(\vt^1 - \vt^2) \nonumber \\
         & (\vt^1,-1+\vartheta^2,0,0) & 1 & \oh(\vt^1 - \vt^2)\nonumber 
\vspace{2mm}\\
{\rm R}  & (-\oh+\vartheta^1,-\oh+\vartheta^2,-\oh,+\oh) &
e^{2\pi i\frac {1}{N}} & 0\\
         &  (-\oh+\vartheta^1,-\oh+\vartheta^2,+\oh,-\oh) &
e^{-2\pi i\frac{1}{N}} & 0
\end{array}
\label{sector5ab}
\eeq}
where $\vt^i \equiv \vt^i_{ab}$ and we have supposed 
 $0 < \vt^i < 1$, $i = 1,2$. In any case, one of the NS states will
be necessarily tachyonic, unless $|\vt^1| = |\vt^2|$ and
both are massless. The spectrum can be found again by projecting
out non-invariant states. In this case, however, we must 
also consider the intersection number of both branes
\beq
I_{ab} \equiv [\Pi_a]\cdot[\Pi_b] = I_{ab}^1 I_{ab}^2 
= (n_a^1 m_b^1- m_a^1\ n_b^2)
(n_a^2 m_b^2- m_a^2 n_b^2).
\label{interfive}
\eeq
This number is a topological invariant associated to the
two 2-cycles the branes wrap. Its absolute value counts 
the net number of intersection between such cycles, thus 
telling us how many replicas of (\ref{sector5ab}) are
present, and its sign indicates the chirality of the
fermions living at the intersection \cite{afiru,cim1}.
The final spectrum arising from this sector is thus
{\small\beq
\begin{array}{rl}
{\rm\bf Tachyons} & \quad \sum_{a<b} \sum_{i=1}^N \; I_{ab}\times
(N_a^i,{\ov N}_b^i) \nonumber \\
{\rm\bf Left\; Fermions} & \quad \sum_{a<b} \sum_{i=1}^N \;
I_{ab}\times(N_a^i,{\ov  N}_b^{i+1}) \nonumber \\
{\rm\bf Right\; Fermions} & \quad \sum_{a<b} \sum_{i=1}^N \;  I_{ab}\times
(N_a^i,{\ov  N}_b^{i-1})
\end{array}
\label{espectro5ab}
\eeq}

In the same manner, we can compute the massless and tachyonic 
spectrum arising from the $D5_aD5_{b^*}$ and $D5_aD5_{a^*}$ sectors,
taking account of their respective wrapping numbers and twist 
vectors. The important point to notice is that fermions arising
from $D5_aD5_{b^*}$ will transform as bifundamentals in some
$(N_a^i,N_b^{-i-1})$ instead of $(N_a^i,{\ov  N}_b^{i+1})$.
This simple fact will allow us to achieve a much more economical
chiral spectrum, as already noted in \cite{imr}. The $D5_aD5_{a^*}$
sector, in turn, will generically have some fixed points under 
the orientifold action, giving rise to fermions and scalars 
in symmetric (${\bf S}$) and antisymmetric (${\bf A}$) 
representations. The complete spectrum is given by
{\small\beq
\begin{array}{l}
{\rm\bf Complex\; Scalars} \\ \sum_{a<b} \sum_{i=1}^N \; 
[\; \arr I_{ab}\arr (N_a^i,{\ov N}_b^i) + 
\arr I_{ab^*}\arr (N_a^i,N_b^{-i})\;]\\
\sum_a [\;2 \arr m_a^1 m_a^2\arr (\arr n_a^1 n_a^2\arr + 1) ({\bf A}_a^0)
+ 2 \arr m_a^1 m_a^2\arr (\arr n_a^1 n_a^2\arr - 1) ({\bf S}_a^0)\;] 
\vspace{3mm}\\
{\rm\bf Left\; Fermions} \\ \sum_{a<b} \sum_{i=1}^N \;
[\;I_{ab}(N_a^i,{\ov  N}_b^{i+1}) + I_{ab*}(N_a^i,N_b^{-i-1})\;]  \\
\sum_a \sum_{j,i=1}^N \; \d_{j,-i-1}
[\;2 m_a^1 m_a^2 (n_a^1 n_a^2 + 1) ({\bf A}_a^j)
+ 2 m_a^1 m_a^2 (n_a^1 n_a^2 - 1) ({\bf S}_a^j)\;] 
\vspace{3mm}\\
{\rm\bf Right\; Fermions} \\ \sum_{a<b} \sum_{i=1}^N \;
[\;I_{ab}(N_a^i,{\ov  N}_b^{i-1}) + I_{ab*}(N_a^i,N_b^{-i+1})\;]  \\
\sum_a \sum_{j,i=1}^N \; \d_{j,-i+1}
[\;2 m_a^1 m_a^2 (n_a^1 n_a^2 + 1) ({\bf A}_a^j)
+ 2 m_a^1 m_a^2 (n_a^1 n_a^2 - 1) ({\bf S}_a^j)\;] 
\end{array}
\label{espectro5ab*}
\eeq}

Let us also mention that, in case $I_{aa*} = 0$, some care 
should be taken when considering the {\it scalar} spectrum
arising from the $aa^*$ sector. If, for instance,
$I_{aa^*}^i = 0$, in order to obtain 
such spectrum we must `forget' about this ${\bf (T^2)_i}$ and 
compute it from a system of D4-branes wrapping as
$(n_a^j, m_a^j)$ on ${\bf (T^2)_j}$, $j \neq i$ 
(see formulae (\ref{espectro4ab*})).
Notice that if $m_a^i = 0$ there is an extra contribution
to the mass$^2$ of the whole spectrum arising from $aa^*$, coming from the 
separation $Y$ that both mirror branes may have in the 
$i^{th}$ torus.

\subsection{Tadpoles and anomalies}

When dealing with a full consistent configuration of D5-branes,
RR tadpole cancellation conditions should always be satisfied.
These can be easily computed from usual factorization of
one-loop amplitudes. As mentioned before, the presence 
of the $\OR$ factor will induce non-vanishing
Klein bottle and Moebius strip contributions to such 
amplitudes, so the conditions computed in \cite{afiru}
for D5-branes sitting on an orbifold singularity 
will be slightly modified to
{\beq
\begin{array}{l}
c_k^2 \ \sum_a n_a^1 n_a^2 \ 
\left({\rm Tr} \gamma_{k,a} + {\rm Tr} \gamma_{k,a^*} \right)
= 16 \ {\rm sin} \left(\frac{\pi k}{N} \right) \\
c_k^2 \ \sum_a m_a^1 m_a^2 \ 
\left({\rm Tr} \gamma_{k,a} + {\rm Tr} \gamma_{k,a^*} \right) = 0\\
c_k^2 \ \sum_a n_a^1 m_a^2 \ 
\left({\rm Tr} \gamma_{k,a} - {\rm Tr} \gamma_{k,a^*} \right) = 0\\
c_k^2 \ \sum_a m_a^1 n_a^2 \ 
\left({\rm Tr} \gamma_{k,a} - {\rm Tr} \gamma_{k,a^*} \right) = 0
\label{tadpoleO5b}
\end{array}
\eeq}
where $c_k^2 = {\rm sin \ } \frac{2\pi k}{N}$ is a weight for each
$k^{th}$ twisted sector usually arising in $\ent_N$ orientifold 
compactifications \cite{GJ}. As can easily be seen,
the difference with the orbifold case amounts to consider the presence 
of mirror branes $a^*$ in our configuration and including a constant
term in the first equation. This constant term can be interpreted as
a negative RR charge induced by the presence of an O5-plane. Indeed,
in the more general context of D$5_a$-branes wrapping general 2-cycles 
$[\Pi_a]$ on ${\bf T^4}$ these conditions can be expressed as
\beq
c_k^2 \ \sum_a \left([\Pi_a] \ {\rm Tr} \gamma_{k,a} 
+ [\Pi_{a^*}] \ {\rm Tr} \gamma_{k,a^*} \right)
= [\Pi_{O5}] \ 16 \b^1\b^2 \
{\rm sin} \left(\frac{\pi k}{N} \right),
\label{tadpoleO5}
\eeq
where $[\Pi_{O5}]$ describes the 2-cycle the O5-plane wraps,
and $\b^i = 1 - b^{(i)}$. Notice that the factor of 
$16 \b^1\b^2$ can be interpreted as the number of O$5$-planes,
which is $4 \b^1\b^2$, times their relative charge to a 
D$5$-brane, which is $-4$. We thus see that RR conditions
can be interpreted, as usual, as the vanishing of the total
RR charge in a compact space (in our case ${\bf T^4}$). 
In this token, notice that $c_0^2 = 0$, so we are not imposing 
any condition in the untwisted sector, whose associated RR form
can escape the singularity. This implies that we are not fixing
the total number of branes. In this sense, we are being less
restrictive than in a simple toroidal orbifold 
(see, e.g., the related  constructions considered in \cite{honecker}).
When embedding our orientifold singularity in a 
full compact variety $\cam$, however, these RR untwisted conditions
should also be taken into account. The cancellation of these untwisted
tadpoles is easy to achieve by adding appropriate D5-branes
at locations in the third torus away from the $\ent_N$ singularity
at which the SM branes sit. This is why we will not discuss them
explicitly in the rest of the paper.

Although quite general, the expression (\ref{tadpoleO5}) is not 
very useful for our model-building purposes. We will make use
instead of (\ref{tadpoleO5b}), which can be also be converted into a 
more tractable expression. Indeed, notice that the upper 
set of equations in (\ref{tadpoleO5b}) is equivalent to
\beq
\sum_a{n_a^1 n_a^2 \left({\rm Tr} \gamma_{2k,a} 
+ {\rm Tr} \gamma_{2k,a^*}\right)} =
{16 \over {\alpha^k + \alpha^{-k}}},
\label{decomp}
\eeq
where we have again used $\alpha = e^{2\pi i/N}$.
Taking $2k \equiv 1  \ mod \ N$, we can easily read
the condition that has to be imposed to the
Chan-Paton matrix $\gamma_{\om,a}$
\beq
\sum_a{n_a^1 n_a^2 \left({\rm Tr} \gamma_{\om,a} 
+ {\rm Tr} \gamma_{\om,a^*}\right)} =
{16 \over {\alpha^{N+1 \over 2}  + \alpha^{N-1 \over 2}}} =
16 \eta \sum_{l=1}^r{(\alpha^{2l-1}+ \bar\alpha^{2l-1})},
\label{generadortwist}
\eeq
\beq
\eta = \left\{\begin{array}{l}
+1 \ {\rm if} \ N = 4r-1 \\
-1 \ {\rm if} \ N = 4r+1
\end{array}\right.
\label{eta}
\eeq

Cancellation of RR tadpoles has, as usual, very important
consequences from the point of view of the effective
four-dimensional field theory. Indeed, when considering
a chiral spectrum as the one considered in (\ref{espectro5ab*})
potential chiral anomalies may arise. RR tadpole conditions
(\ref{tadpoleO5b}) insure the cancellation of such anomalies,
as we will now see. Let us first consider the cancellation of 
the cubic non-Abelian anomaly for the gauge group $SU(N_a^i)$,
which in our configurations reads
\beq
\ca_{SU(N_a^i)^3} = \sum_{b,j} N_b^j \left(I_{ab} \ \d(i,j) 
+ I_{ab^*} \ \d(i,-j)\right) + 16\b^1\b^2 \ I_{a,O5}\ \d(i,-i), 
\label{quiralO5}
\eeq
where $\d(i,j) = \d_{i+1,j} - \d_{i-1,j}$ 
(the indexes $i,j$ are again defined $mod$ $N$) 
and, in case of factorizable branes (\ref{brana_a}), 
$\b^1\b^2 I_{a,O5} = m_a^1 m_a^2$.

Just as done in \cite{leigh,afiru}, we can use the discrete
Fourier transform 
$\d_{ij} = \frac 1N \sum_{k=1}^N e^{\frac{2\pi i k}{N} (j-i)}$ 
to rewrite (\ref{quiralO5}) as 
\beq
\ca_{SU(N_a^i)^3} = {-4 \over N} \sum_{k=1}^N e^{2\pi i\frac{k\cdot i}{N}} 
c_k^2 \left( \sum_b I_{ab} \ {\rm Tr \ }\g_{k,a}\ 
+ \ I_{ab^*} {\rm Tr \ }\g_{k,a^*}\right) + 16 m_a^1 m_a^2 \d(i,-i),
\label{quiralO5b}
\eeq
which after some simple manipulations, can be seen to vanish whenever
the tadpoles conditions (\ref{tadpoleO5b}) are satisfied. As usual,
the latter turn out to be more restrictive that the vanishing of
(\ref{quiralO5b}).

We can also consider mixed and cubic $U(1)$ anomalies, both involving 
a generalized Green-Schwarz mechanism mediated by RR twisted fields.
Indeed, the full expression for the mixed $U(1)_{a,i}-SU(N_b^j)^2$
anomaly is given by
\beqa
\ca_{U(1)_{a,i}-SU(N_b^j)^2} & = &\oh \d_{ab} \d_{ij} 
\left(
\sum_{c,l} N_c^l \left[ I_{ac}\ \d(i,l) + I_{ac^*}\ \d(i,-l) \right]
+ 16 m_a^1 m_a^2 \d(i,-i)
\right) \nonumber \\
& + & \oh N_a^i \left( I_{ab}\ \d(i,j) + I_{ab^*}\ \d(i,-j) \right),
\label{mixtaO5}
\eeqa
the first term in brackets being proportional to the cubic
chiral anomaly of $SU(N_a^i)$, thus vanishing when imposing 
tadpoles. The remaining contribution can then be canceled by
means of a generalized Green-Schwarz mechanism. 
Indeed, by use of the discrete Fourier
transform, we can rewrite the residual anomaly in (\ref{mixtaO5})
as
\beq
\ca_{U(1)_{a,i}-SU(N_b^j)^2} = {-2 N_a^i \over N} 
\sum_{k=1}^N e^{2\pi i\frac{k\cdot i}{N}} 
c_k^2 \left( I_{ab} \ e^{-2\pi i\frac{k\cdot j}{N}}
+ \ I_{ab^*}  \ e^{2\pi i\frac{k\cdot j}{N}} \right).
\label{mixtaO5b}
\eeq
As explained in \cite{afiru} for the orbifold (non-orientifold) 
case, this quantity can be canceled by exchange of four-dimensional 
fields, which arise upon dimensional reduction of the
RR twisted forms living on the singularity. 
For showing this, let us consider the T-dual picture of
fractional D7-branes wrapping the first two tori, and with
non-trivial $F$ and $B$-fluxes on them. 
On the worldvolume of each D7-brane, there will appear some 
couplings to the twisted RR forms in (\ref{pformas}),
and by integrating such couplings on the compact toroidal dimensions
${\bf (T^2)_1 \times (T^2)_2}$ we will obtain four-dimensional
couplings that will be relevant to our low-energy theory.
Indeed, if we define
\beq
\begin{array}{lcl}
B_0^{(k)} = A_0^{(k)}, 
& & B_2^{(k)} = \int_{\bf (T^2)_1 \times (T^2)_2} A_6^{(k)}, \\
C_0^{(k)} = \int_{\bf (T^2)_1 \times (T^2)_2} A_4^{(k)}, 
& & C_2^{(k)} =  A_2^{(k)}, \\
D_0^{(k)} = \int_{\bf (T^2)_2} A_2^{(k)},
& & D_2^{(k)} = \int_{\bf (T^2)_1} A_4^{(k)}, \\
E_0^{(k)} = \int_{\bf (T^2)_1} A_2^{(k)}, 
& & E_2^{(k)} = \int_{\bf (T^2)_2} A_4^{(k)},
\end{array}
\label{pformas4}
\eeq
then these four dimensional couplings can be computed to be
\beqa & &
\begin{array}{c}
c_k N_a^i\, n^1_a n^2_a \int_{M_4} 
{\rm Tr} \left(\g_{k,a}-\g_{k,a^*}\right)\lam_i \
B_2^{(k)}\wedge {\rm Tr} F_{a,i}, \\
c_k N_a^i\, m^1_a m^2_a \int_{M_4} 
{\rm Tr} \left(\g_{k,a}-\g_{k,a^*}\right)\lam_i \
C_2^{(k)}\wedge {\rm Tr} F_{a,i}, \\
c_k N_a^i\, n^1_a m^2_a \int_{M_4} 
{\rm Tr} \left(\g_{k,a}+\g_{k,a^*}\right)\lam_i \
D_2^{(k)}\wedge {\rm Tr} F_{a,i}, \\
c_k N_a^i\, m^1_a n^2_a \int_{M_4} 
{\rm Tr} \left(\g_{k,a}+\g_{k,a^*}\right)\lam_i \
E_2^{(k)}\wedge {\rm Tr} F_{a,i},
\end{array} 
\label{acoplosdualO5} \\ & &
\begin{array}{c}
c_k m^1_b m^2_b \int_{M_4} 
{\rm Tr} \left(\g_{k,b}^{-1}+\g_{k,b^*}^{-1}\right)\lam_j^2 \
B_0^{(k)} \wedge {\rm Tr} \left(F_{b,j}\wedge F_{b,j}\right), \\
c_k n^1_b n^2_b \int_{M_4} 
{\rm Tr} \left(\g_{k,b}^{-1}+\g_{k,b^*}^{-1}\right)\lam_j^2 \
C_0^{(k)} \wedge {\rm Tr} \left(F_{b,j}\wedge F_{b,j}\right), \\
c_k m^1_b n^2_b \int_{M_4} 
{\rm Tr} \left(\g_{k,b}^{-1}-\g_{k,b^*}^{-1}\right)\lam_j^2 \
D_0^{(k)} \wedge {\rm Tr} \left(F_{b,j}\wedge F_{b,j}\right), \\
c_k n^1_b m^2_b \int_{M_4} 
{\rm Tr} \left(\g_{k,b}^{-1}-\g_{k,b^*}^{-1}\right)\lam_j^2 \
E_0^{(k)} \wedge {\rm Tr} \left(F_{b,j}\wedge F_{b,j}\right),
\end{array}
\label{acoplosdual2O5}
\eeqa
where $\lam$ denotes the Chan-Paton wavefunction for the gauge 
boson state, and the $N_a^i$ factor arises from normalization of 
the $U(1)_{a,i}$ generator (see \cite{iru}). 
 Since $B_2^{(k)}$ and $B_0^{(k)}$ are
four-dimensional Hodge duals, same for $C$, $D$ and $E$,
the sum over the GS diagrams
will provide a counterterm
with the structure (\ref{mixtaO5b}), just as required 
to cancel the residual mixed anomaly in (\ref{mixtaO5}).
Cancellation of cubic $U(1)$ anomalies works in a similar way.

An important consequence of this anomaly cancellation mechanism
is the Abelian gauge structure of the low-energy
effective action. It can be shown that, as a result of
the couplings (\ref{acoplosdualO5})
the gauge bosons of the potentially anomalous 
$U(1)$ get massive, decoupling from the low energy spectrum
of the theory. More generally,
any $U(1)$ gauge boson (anomalous or not)  
with a non-vanishing axionic coupling
of the form (\ref{acoplosdualO5}) will have an induced mass term
of the order of the string scale. The associated gauge symmetry
will no longer be present, although such $U(1)$ will remain
as an exact perturbative global symmetry.

A similar analysis regarding the construction of intersecting
D$4$-branes configurations wrapping ${\bf T^2}$ and sitting in a 
$\C^2/\ent_N$  orientifold singularity can also be performed, 
the general formalism being much alike as the one just presented for
the case of D5-branes. These D4-branes constructions are
also of  interest from the model-building point of view,
and some non-orientifold examples were built in \cite{afiru2}
(see also  related  models 
in \cite{honecker}). However, 
it turns out to be difficult to obtain D4-brane
models with just the SM fermion spectrum.
That is why  we leave the presentation of the
D4-brane formalism for an appendix.

\section{The Standard Model at intersecting D5-branes }

In the present section we will be interested in 
 finding intersecting
D5-branes models whose gauge group and matter content correspond
to either the Standard Model (SM) or some Left-Right symmetric (LR) 
extension of it \cite{LR}. Such low energy spectra must contain 
the following gauge group and fermionic content:
\beq
\begin{array}{ccc}
{\rm Standard\ Model} &  & {\rm Left}$-${\rm Right\ Model}
\vspace{0.1cm} \\ 
SU(3)_c \ti SU(2)_L \ti U(1)_Y 
& & SU(3)_c \ti SU(2)_L \ti SU(2)_R \ti U(1)_{B-L} \\
Q_L^i=(3,2)_{\frac 16} & \raw & Q_L^i=(3,2,1)_{1/3}\\ 
\left.
\begin{array}{c} 
U_R^i =({\bar 3},1)_{-\frac 23} \\
D_R^i=({\bar 3},1)_{\frac 13} 
\end{array} \right\}
& \raw & Q_R^i=(\bar 3,2,1)_{-1/3} \\
L^i=(1,2)_{-\frac 12} & \raw & L_L^i=(1,2,1)_{-1}  \\ 
\left.
\begin{array}{c} 
E_R^i=(1,1)_1 \\
N_R^i=(1,1)_0
\end{array} \right\}
& \raw & L_R^i=(1,1,2)_{1}
\end{array}
\label{content}
\eeq
where $i = 1,2,3$ indexes the three different generations 
that have to be considered in each model. 

Following the general philosophy described in \cite{imr},
we will be considering a class of configurations where
the chiral fermions arise only in bifundamental representations
\beq
\sum_{a,b}
n_{ab}(N_a,{\overline N}_b)+m_{ab}  (N_a,N_b) + n_{ab}^* ({\overline N}_a,
N_b)+m_{ab}^*({\overline N}_a , {\overline N}_b),
\label{bifundamentals}
\eeq
where  $n_{ab},n_{ab}^*,m_{ab},m_{ab}^*$ are model dependent 
and non-negative integer numbers. In this particular class of 
models, cubic anomaly cancellation for a non-Abelian gauge group 
$SU(N_a)$ reduces to having the same 
number of fundamental representations $N_a$ as antifundamental
representations $\overline N_a$. 
Notice also that, from the
point of view of Left-Right unification, right-handed neutrinos 
must exist, as they complete the $SU(2)_R$ leptonic doublet 
that contains the charged right-handed leptons $E_R^i$. 
From the point of view of SM building, though, there is no reason
why we should consider having such representations in our fermionic 
content. However, as was emphasized in \cite{imr}, when obtaining
the chiral content of our theory just from fields transforming in 
bifundamental representations, such right-handed neutrinos naturally
appear from anomaly cancellation conditions. Since in the present
paper we will construct our models from such ``bifundamental'' fermions, 
we will include these particles right from the start
\footnote{For some intersecting branes SM constructions without 
right-handed neutrinos see \cite{bklo,cim2}.}.
In general, it can also be shown that 
in this case where chiral fields transform in bifundamentals
the simplest embedding of the SM (or the LR extension)
will consist in a configuration of four stack of branes, as
presented in table \ref{SMbranes}.
\begin{table}[htb]
\renewcommand{\arraystretch}{1.7}
\begin{center}
\begin{tabular}{|c|c|c|c|}
\hline
Label & Multiplicity & Gauge Group & Name \\
\hline
\hline
stack $a$ & $N_a = 3$ & $SU(3) \times U(1)_a$ & Baryonic brane\\
\hline
stack $b$ & $N_b = 2$ & $SU(2)_L \times U(1)_b$ & Left brane\\
\hline
stack $c$ & $N_c = 
\left\{
\begin{array}{c}
2\\ 1
\end{array}
\right.$ & 
$\begin{array}{c}
SU(2)_R \ti U(1)_c \\ U(1)_c
\end{array}$ & Right brane\\
\hline
stack $d$ & $N_d = 1$ & $U(1)_d$ & Leptonic brane \\
\hline
\end{tabular}
\caption{\small Brane content yielding the SM or LR spectrum.
\label{SMbranes}}
\end{center}
\end{table}

Given this brane content is relatively easy to figure out
how to realize the specific fermion content of both SM and LR
models. Indeed, let us for instance consider strings coming 
from the $ab$ and $ab^*$ sectors. Their (left-handed) massless modes will 
transform as either $(3,\bar 2)$ or $(3,2)$ under the gauge 
group $SU(3) \ti SU(2)_L$
and hence can be naturally identified 
with the left-handed quarks $Q_L^i$. The fermion content  
 of both classes of models are shown in tables
\ref{specSM} and \ref{specLR}, where each chiral fermion in 
(\ref{content}) is associated to a definite sector.

\begin{table}[htb] \footnotesize
\renewcommand{\arraystretch}{1.25}
\begin{center}
\begin{tabular}{|c|c|c|c|c|c|c|c|}
\hline Intersection &
 Matter fields  &   &  $Q_a$  & $Q_b $ & $Q_c $ & $Q_d$  & Y \\
\hline\hline (ab) & $Q_L$ &  $(3,2)$ & 1  & -1 & 0 & 0 & 1/6 \\
\hline (ab*) & $q_L$   &  $2(3,2)$ &  1  & 1  & 0  & 0  & 1/6 \\
\hline (ac) & $U_R$   &  $3( {\bar 3},1)$ &  -1  & 0  & 1  & 0 & -2/3 \\ 
\hline (ac*) & $D_R$   &  $3( {\bar 3},1)$ &  -1  & 0  & -1  & 0 & 1/3 \\
\hline (bd) & $L$    &  $3(1,2)$ &  0   & -1   & 0  & 1 & -1/2  \\
\hline (cd) & $N_R$   &  $3(1,1)$ &  0  & 0  & 1  & -1  & 0  \\
\hline (cd*) & $E_R$   &  $3(1,1)$ &  0 & 0 & -1 & -1  & 1 \\
\hline \end{tabular}
\end{center} \caption{\small Standard model spectrum and $U(1)$ charges. 
The hypercharge generator is defined as 
$Q_Y = \frac 16 Q_a - \frac 12 Q_c - \frac 12 Q_d$.}
\label{specSM} 
\end{table}
\begin{table}[htb] \footnotesize
\renewcommand{\arraystretch}{1.25}
\begin{center}
\begin{tabular}{|c|c|c|c|c|c|c|c|}
\hline Intersection &
 Matter fields  &   &  $Q_a$  & $Q_b $ & $Q_c $ & $Q_d$  & $B-L$ \\
\hline\hline (ab) & $Q_L$ &  $(3,2,1)$ & 1  & -1 & 0 & 0 & 1/3 \\
\hline (ab*) & $q_L$   &  $2(3,2,1)$ &  1  & 1  & 0  & 0  & 1/3 \\
\hline (ac) & $Q_R$   &  $({\bar 3},1,2)$ &  -1  & 0  & 1  & 0 & -1/3 \\
\hline (ac*) & $q_R$   &  $2({\bar 3},1,2)$ &  -1  & 0  & -1  & 0 & -1/3 \\
\hline (bd) & $L_L$    &  $3(1,2,1)$ &  0  & -1  & 0  & 1 & -1  \\
\hline (cd) & $L_R$   &  $3(1,1,2)$ &  0  & 0  & 1  & -1  & 1  \\
\hline \end{tabular}
\end{center} \caption{\small Left Right symmetric chiral spectrum
and $U(1)$ charges. The $U(1)_{B-L}$ generator is defined as 
$Q_{B-L} = \frac 13 Q_a - Q_d$.}
\label{specLR}
\end{table}

\begin{figure}
\centering
\epsfxsize=2.5in
\hspace*{0in}\vspace*{.2in}
\epsffile{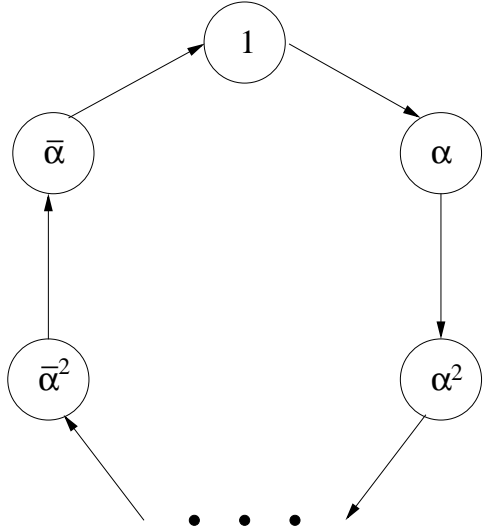}
\caption{\small{Quiver diagram of a $\ent_N$ orbifold singularity.
The nodes of such diagram represent the phases associated to each 
different gauge group in the theory, whereas each arrow
represents a chiral fermion transforming in a bifundamental
of the two groups it links.}}
\label{quiverZN}
\end{figure}

In order to realize such spectra as the chiral content of a
concrete configuration of D5-branes we must impose some topological
constraints on  our models. Unlike the case of D6-branes, where all 
the spectrum information is encoded on the intersection numbers,
we must now also consider the orbifold structure of our configuration.
Such structure can be easily encoded in a quiver diagram
\footnote{These are quivers in the sense of ref.\cite{dg,quiver,auquivers},
not in the sense of the SUSY-quivers discussed in ref.\cite{cim1,cim2}
in which no $\ent_N$ twist is present.},
as shown in figure \ref{quiverZN}. 

In general, a D-brane configuration living on an orbifold 
singularity can be locally described by quotienting the theory 
by a discrete group $\Gamma$, which is acting 
on both an ambient space $\C^n$ and on the CP 
degrees of freedom. To each $\Gamma$ action we can associate 
a quiver diagram \cite{dg,quiver,auquivers}. Each node of such diagram 
will represent an irreducible representation (irreps) of $\Gamma$,
whereas the arrows connecting the nodes represent invariant fields
under combined geometric and gauge actions. In general, the $\G$
action $\g_g$ on the Chan-Paton degrees of freedom can be written
as a direct sum of such irreps, and the gauge groups that will arise
from it will correspond to a product of unitary groups, each one
associated to a definite irreps. In our specific setup
$n = 1$ and $\Gamma = \ent_N$, so each irreps of $\G$ is one-dimensional
and can be associated to a $N^{th}$-root of unity. Indeed,
any $\ent_N$ generator action on the Chan-Paton degrees of freedom can be
written on the form (\ref{Chan}), where several such phases are involved.
Without loss of generality, we will consider that each brane
$a, b, c, d$ has a $\g_\om$ matrix proportional to the identity, 
that is $\g_{\om,i} = \a^n {\bf 1}_{N_i}$, so that
it will give rise to just one unitary gauge group $U(N_i)$. 
We will represent this by locating that brane $i$ on the node
corresponding to the irreps $\a^n$. Notice that, in an orientifold
theory, the mirror brane $i^*$ will then be placed in the node
$\bar \a^n$.

Chiral fields can also be easily identified in this diagram 
by arrows connecting the nodes. These arrows will always link 
two different nodes, so that if there is some brane content 
in both of them we will find a fermion transforming under
the corresponding gauge groups. The sense of the arrow will
denote the chirality that such representation has.
In our conventions the positive sense represents left
fermions. This arrow structure can be easily read from the chiral 
spectrum in (\ref{espectro5ab}), giving rise to the cyclic quiver
depicted in figure \ref{quiverZN}. Notice that this simple spectrum
comes from a plain orbifold singularity. In this case every 
chiral field will transform in bifundamental representations of
two gauge groups with contiguous phases. When considering
orientifold singularities, however, we should also include the mirror
branes on the picture, and more ``exotic'' representations may arise.

There are, in principle, many different ways of obtaining the
desired chiral spectrum (\ref{content}) from the brane content
of table \ref{SMbranes}. Furthermore, the details of the 
construction will depend on the specific model (SM or LR)
and on the $\ent_N$ quiver under consideration.
There are, however, some general features of the
construction that can be already addressed at this level.

\begin{itemize}

\item 
In both SM and LR models, chiral fermions must arise in a very
definite pattern. Namely, we need left and right-handed quarks ,
so we must consider matter arising from the intersections
of the {\it baryonic} brane with both the {\it left} and 
{\it right} branes. We must avoid, however, lepto-quarks
which may arise from some intersection with the {\it leptonic}
brane. The same considerations must be applied to the latter.
This pattern can be easily achieved in D5-branes configurations
by placing both $b$ and $c$ (or $c^*$) branes on the same node of the 
$\ent_N$ quiver, while $a$ and $d$ in some contiguous node.
Since, in order to achieve the spectra of tables 
\ref{specSM} and \ref{specLR},
we must consider non-trivial $ab$, $ab^*$, $ac$ and $ac^*$
sectors, we must place the stack $a$ either in the phase
$1$ or in the phase $\a$, while stacks $b$, $c$ must be in 
the other one. This restricts our search to essentially 
two different distributions of branes, which are shown in 
figure \ref{2quivers}.

\begin{figure}
\centering
\epsfxsize=4in
\hspace*{0in}\vspace*{.2in}
\epsffile{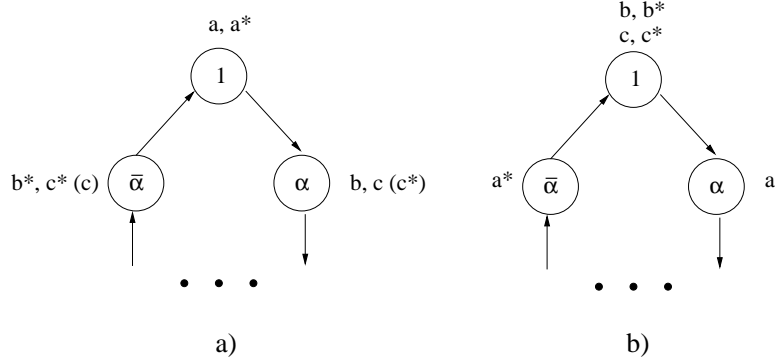}
\caption{\small{Two possible embeddings of the brane content
of a SM or LR configuration.}}
\label{2quivers}
\end{figure}

\item 
Given these two possibilities, it is now easy to guess which
intersection numbers must we impose in order to achieve
the desired spectra. Indeed, the modulus of an intersection 
number, say $I_{ab}$, will give us the multiplicity of
this sector. This implies that, in order to have the desired
number of left-handed quarks, we must impose $|I_{ab}| = 1$,
$|I_{ab^*}| = 2$ as can be read directly from tables 
\ref{specSM} and \ref{specLR}.
\footnote{We could have alternatively imposed $|I_{ab}| = 2$,
$|I_{ab^*}| = 1$, giving an equivalent spectrum.
}
On the other hand, we will have to choose the sign of these
intersection numbers in order to properly fix the chirality
of our fermions. These signs will be different for each
distribution of branes considered in figure \ref{2quivers}, 
since chirality also depends on the arrow structure of the 
quiver diagram. For instance, we should impose 
$I_{ab} = 1$, $I_{ab*} = -2$ in the $a$)-type of quiver
in this figure, while $I_{ab} = -1$, $I_{ab*} = -2$
in the $b$)-type. Similar reasoning  applies  to  other intersections
involving branes $b$ and $c$.

\item 
Finally, we are interested in getting all of our chiral matter 
from bifundamental representations. Thus, we must avoid the 
appearance of Symmetrics and Antisymmetrics that might appear 
from the general spectrum (\ref{espectro5ab*}).
This will specially arise in $\ent_3$ models, where
we will have to impose $I_{ii^*} = 0$ for those branes 
in the $\a$ node. 

\end{itemize}

\subsection{D5 Standard Models}

\begin{figure}
\centering
\epsfxsize=4.5in
\hspace*{0in}\vspace*{.2in}
\epsffile{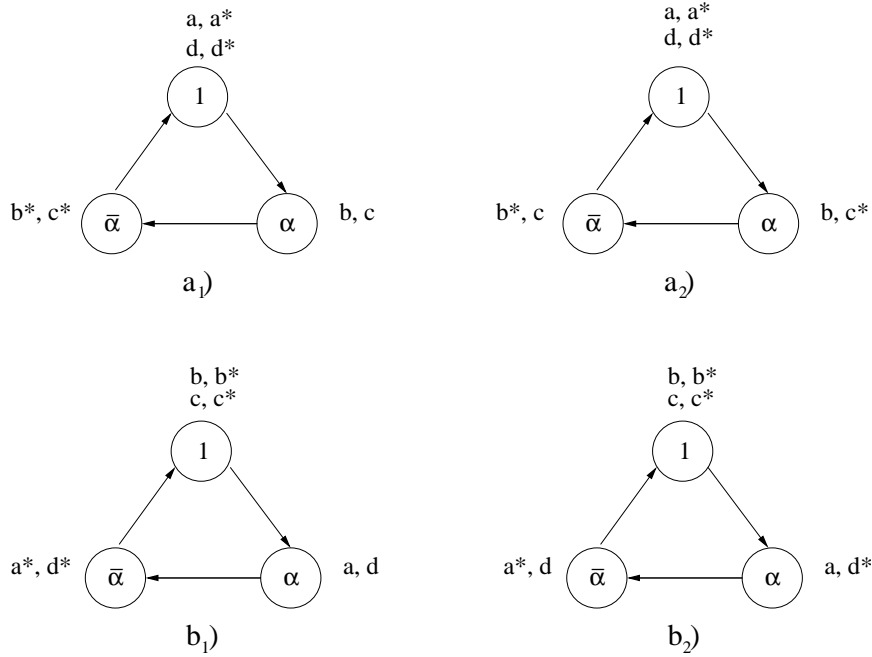}
\caption{\small{Four possible embeddings of the brane content
of a SM configuration in a $\ent_3$ quiver.}}
\label{SMfig}
\end{figure}

Let us give an example that shows how the SM structure can be
implemented on D5-branes configurations.
The simplest choice for such example is the $\ent_3$ singularity, 
which is the smallest $\ent_N$ quiver that provides 
non vector-like spectra. 
Imposing the chirality pattern discussed above give us four different
ways of embedding the SM spectrum, each of them depicted in figure 
\ref{SMfig}.
In order to achieve a SM configuration, we must
impose the intersection numbers that will give us the 
desired matter content. As discussed above, these will depend on
the particular $\ent_3$ quiver considered. Let us first consider the quiver
$a_1$). In table \ref{SMatab} we show the general class of solutions
for the wrapping numbers that will provide us with such fermionic spectrum.

\begin{table}[htb] 
\footnotesize
\renewcommand{\arraystretch}{1.7}
\begin{center}

\begin{tabular}{|c||c|c|c|}
\hline
 $N_i$    &  $(n_i^1,m_i^1)$  &  $(n_i^2,m_i^2)$   & $\gamma_{\om,i}$ \\
\hline\hline 
$N_a = 3$ & $(n_a^1, \ep \b^1)$  &  $(3,- \oh \ep \tilde \ep)$ & 
$1_3$  \\
\hline 
$N_b = 2$ & $(1/\b^1, 0)$ &
$(\tilde \ep, -\oh \ep)$ &  $\a 1_2$   \\
\hline 
$N_c = 1$ & $(1/\b^1, 0)$ & $(0, \ep)$ & $\a$ \\
\hline 
$N_d=1$ & $(n_d^1, 3\ep \b^1)$ & $(1, \oh\ep\tilde\ep) $ & $1$ \\
\hline
$N_h$ & $(\ep_h/\b^1, 0)$ & $(2,0)$ & $1_{N_h}$  \\
\hline
\end{tabular}

\caption{\small{D5-branes wrapping numbers and CP phases giving rise
to a SM spectrum the $\ent_3$ quiver of fig \ref{SMfig}.$a_1$). The solution
is parametrized by $n_a^1, n_d^1 \in \ent$, $\ep, \tilde\ep = \pm 1$
and $\b^1 = 1 - b^{(1)} = 1, 1/2$. Notice that the second torus has
to be tilted, hence $\b^2 = 1/2$.}
\label{SMatab}}
\end{center}
\end{table}

Notice that for the sake of generality  
we have added a new stack of $N_h$ branes to our initial 
configuration, yielding an extra $U(N_h)$ gauge group. However,
the wrapping numbers and the CP phase of such brane have
been chosen in such a way that no extra chiral matter arises from
its presence. Since no chiral fermion is charged under the gauge group
of this brane, the stack $h$ is a sort of hidden sector of the 
theory. This is strictly true, however, only from the fermion 
content point of view, and generically some scalars 
with both SM and $U(N_h)$ quantum numbers may  appear.

Having achieved the fermionic spectrum of table \ref{specSM},
our low energy field theory will be automatically free of 
cubic chiral anomalies. In order to have a consistent compactification, 
however, we must impose the stronger tadpole cancellation conditions.
Interestingly enough, most of the conditions in (\ref{tadpoleO5b})
turn out to be trivially satisfied by this brane content, the only
non-trivial one being the first condition, that now reads
\beq
9n_a^1 + n_d^1 - \frac{\tilde\ep}{\b^1} + 2N_h \frac{\ep_h}{\b^1}
= -8.
\label{tadSMZ3a}
\eeq

Let us now analyze the $U(1)$ structure of such model. As described in
the previous section, couplings of gauge bosons to
twisted RR fields will give rise to GS counterterms that will cancel 
the residual $U(1)$ anomalies. We are particularly interested in
couplings (\ref{acoplosdualO5}), that tell us which gauge bosons
are becoming massive by this mechanism. In the $\ent_3$ orientifold
case there is only one independent twisted sector, so only four
couplings are relevant. By considering the brane content
above we find that these couplings are
\beq
\begin{array}{rcl}
B_2^{(1)} & \wedge  & \ c_1\ (\a -\a^2) {{2\tilde\ep}\over {\beta^1}}
F^{b} \\
D_2^{(1)} & \wedge  & \ c_1 
\left( \ep \tilde\ep (-3n_a^1 F^a\ +\ n_d^1 F^d)\ +\
\frac{\ep}{\b^1}(F^b\ -\ F^c) \right) \\
E_2^{(1)} & \wedge  & \ c_1\ 6\ep\b^1 (3F^a\ +\  F^{d})
\end{array}
\label{bfsSMZ3a}
\eeq
the coupling to the $C_2^{(1)}$ field being trivially null.
In general, such couplings will give mass to three linearly
independent combinations of $U(1)$'s, leaving just one $U(1)$
as a true Abelian gauge symmetry of the spectrum. Among these 
massive $U(1)$'s, two are model-independent, and correspond 
to the `anomalous' combinations $U(1)_b$ and $3U(1)_a+U(1)_d$ 
characteristic of this fermionic spectrum. The third one,
however, will depend on the specific model considered. 
Indeed, we find that the generator of the massless $U(1)$ 
is given by
\beq
Q_0 = Q_a - 3 Q_d - 3\tilde\ep \b^1 (n_a^1 + n_d^1) Q_c,
\label{masslessSMZ3a}
\eeq
so if we further impose to our class of models the condition
\beq
\tilde\ep \b^1 (n_a^1 + n_d^1) = 1,
\label{condSMZ3a}
\eeq
then we find that this
massless Abelian gauge group precisely corresponds to the hypercharge,
which in these models is given by 
$U(1)_Y = \frac 16 U(1)_a - \frac 12 U(1)_c - \frac 12 U(1)_d$.

Notice that the whole of this construction is quite analogous to the 
one described in \cite{imr}. Indeed, we have imposed the same chiral
spectrum, again arising from bifundamental representation of four stack
of branes. After imposing some conditions regarding tadpoles and
the Abelian gauge structure, we are finally led to a compactification 
yielding just the gauge and fermionic spectrum of the Standard Model 
(and possibly  some hidden sector of the theory given by the brane $h$).

\subsubsection{Scalars and tachyons in the spectrum}

As explained in Section 2.1, at the intersection of pairs of
D5-branes with  the {\it same} CP phase there may appear 
scalar tachyons with masses given in eq.(\ref{sector5ab}).
Since branes $b,c$ and their mirrors are parallel along the 
first 2-torus, they generically do not intersect. On the other 
hand there may be tachyons at the intersections $(aa^*),(dd^*)$,
$(ad),(ad^*)$ plus possibly others involving the hidden branes $h$. 
One can get rid of many of these tachyons by appropriately
choosing some discrete parameters and the compactification radii. 
Consider for instance the following choice of parameters:
\beq
n_a^1 = n_d^1\ =\ -1,\ N_h=0,\  \tilde \epsilon\ =\ -1,\ 
\b^1\ =\ \oh   \ .
\label{choice}
\eeq
%
With this choice it is easy to check that the tadpole cancellation 
conditions  (\ref{tadSMZ3a}) are verified and the standard hypercharge 
is the only $U(1)$ remaining at the massless level. 
Furthermore, the $h$ brane is not needed in order to cancel tadpoles,
this hidden sector thus being absent. 
Now, the angles formed by the branes $d,a$ with the orientifold plane on the 
two tori are given by
\beqa
\theta_a^1 = \ep\ \left(\pi - tg^{-1}\left(\frac{U^1}{2}\right) \right) \ 
& ; & 
\theta_a^2 = \ep\  tg^{-1}\left(\frac{U^2}{6}\right) 
\nonumber \\
\theta_d^1 = \ep\ \left(\pi - tg^{-1}\left(\frac{3U^1}{2}\right) \right) \ 
& ; & 
\theta_d^2 = - \ep\  tg^{-1}\left(\frac{U^2}{2}\right)
\label{angulillos}
\eeqa
respectively. Here $U^i=R_2^i/R_1^i$, $i=1,2$. 
Now, the angles formed by such branes with their mirrors 
is given by $\vt_{a,d}^i \equiv - 2 \th_{a,d}^i$ mod $2\pi$,
so for $U^1 = U^2/3$ one gets 
$\arr \vt_{a,d}^1 \arr = \arr \vt_{a,d}^2 \arr$,
and according to eq.(\ref{sector5ab}) the scalars in $(aa^*)$ and $(dd^*)$
cease to be tachyonic and become massless
\footnote{Actually, according to (\ref{espectro5ab*}), scalars in
the sector $(dd^*)$ transform in the antisymmetric representation
of $U(N_d) =  U(1)$, thus being absent for any choice of angles.}
. The only tachyonic scalars in the spectrum persist in the
$ad$ and $ad^*$ intersections which have mass$^2$:
\beq
m^2_{ad}\ =\ m^2_{ad^*}\ = -\ \frac 1\pi 
tg^{-1}\left({{U^2}\over 6}\right)M_s^2 \ .
\label{tachyoncognazo}
\eeq
In table \ref{escalares} we present the lightest scalar spectrum arising 
from branes $a$, $d$ and their mirrors when the particular choice
(\ref{choice}) is made.
\begin{table}[htb] \footnotesize
\renewcommand{\arraystretch}{1.25}
\begin{center}
\begin{tabular}{|c|c|c|}
\hline
 Sector   &  Representation  & $\a'$ mass$^2$ \\
\hline\hline ($aa^*$) & 
$\begin{array}{c}
4\b^1\ (3,1)_{1/3} \\
2\b^1\ (6,1)_{1/3} \\
\end{array}$
  & 0  \\
\hline ($ad$) & $4\b^1$ $(3,1)_{2/3}$ & 
$\pm \frac 1\pi tg^{-1}\left({{U^2}\over 6}\right)$   \\  
\hline ($ad^*$) & $4\b^1$ $(3,1)_{-1/3}$ & 
$\pm \frac 1\pi tg^{-1}\left({{U^2}\over 6}\right)$   \\  
\hline \end{tabular}
\end{center} \caption{\small Lighter scalar excitations arising from the
brane content with phase $1$ in table \ref{SMatab}, under the 
choice of parameters (\ref{choice}).
\label{escalares} }
\end{table}

Note however that all the above scalar masses are tree level results and that,
since the models are non-SUSY,  
there are in general important one-loop contributions to the 
scalar masses. Those will be particularly important for the 
coloured objects like the scalars in $(ad)$, $(ad^*)$ sectors
which are color triplets. Those one-loop corrections may be estimated
from the effective field theory (one gauge boson exchange) and
yield \cite{afiru2}
\beq
\Delta m^2(\mu ) \ =\ \sum_a { {4C_F^a \alpha_a(M_s) }\over {4\pi }} M_s^2
f_a \log(M_s/\mu) \ +\ \Delta M^2_{KK/W}
\label{massloop}
\eeq
where the sum on $a$ runs over the different gauge interactions 
under which the scalar transforms and $C_F^a$
is the eigenvalue of the quadratic Casimir in the fundamental representation.
Here $\Delta M^2_{KK/W}$ denotes further contributions which may
appear from the Kaluza-Klein, winding and string 
 excitations if they are substantially lighter than the
string scale $M_s$. The function $f_a$ is given by
\beq
f_a \ =\  { {2+b_a{{\alpha_a(M_s)}\over {4\pi }} t}\over
{1+b_a{{\alpha_a(M_s)}\over {4\pi }}t } }
\eeq
where $t=2\log(M_s/\mu)$ and $b_a$ are the coefficients of the one-loop
$\beta $-functions. These corrections are positive and may easily overcome 
 the tree level result if $U^2$ is not too large.
This is analogous to the one-loop contribution to squark
masses in the MSSM in which for large gaugino masses
the one-loop contribution clearly dominates over the tree-level
soft masses (see e.g. ref.\cite{mssm} and references therein).
Thus in this class of models, apart from the fermion 
spectrum of the SM, one expects the presence of some extra 
relatively light (of order the electroweak scale) coloured 
scalars.

\subsubsection{Electroweak symmetry breaking}

The Higgs sector in this class of theories is relatively similar to the one 
in the models in \cite{imr}.
Consider in particular the SM 
configuration described in the previous subsections. 
Here, the only light scalar with the quantum numbers of a Higgs
boson lives in the $bc$ sector. Branes 
$b$ and $c$ are parallel in the first torus, but if the 
distance  $X_{bc}$  between the branes in that torus
is set to zero the branes intersect at an angle 
\beq
\pi \vt_{bc}^2 = \ep\tilde \ep \left({{\pi}\over 2}
+ tg^{-1}\left(\frac{U^2}{2}\right)\right)\ ,\
\eeq
and at those intersections complex scalar doublets appear with masses
\beq
m^2_{H^{\pm}} \  =\ {{X_{bc}^2}\over {4\pi } } M_s^2 \  \pm \
{{M_s^2}\over 2} |\vt_{bc}^2| \ ;\
\label{masshiggs}
\eeq
There are in fact two scalar doublets with quantum numbers as in table
\ref{higgsses},
\begin{table}[htb] \footnotesize
\renewcommand{\arraystretch}{1.25}
\begin{center}
\begin{tabular}{|c|c|c|c|}
\hline
 Higgs   &  $Q_b$  &  $Q_c$   & Y \\
\hline\hline $H_1$ & 1  &  -1 & 1/2  \\
\hline $H_2$ & -1  & 1 &  -1/2   \\
\hline \end{tabular}
\end{center} \caption{\small Electroweak Higgs fields
\label{higgsses} }
\end{table}

and defined as
\beq
H^{\pm}={1\over2}(H_1^*\pm H_2) \ .  
\eeq
The intersection number of these branes in the second torus is 
equal to $\pm 1$ so that only one copy of this Higgs system appears.
Thus in the present model we have the same minimal Higgs sector
as in the MSSM.
As may be seen from eq.(\ref{masshiggs}) as the distance
$X_{bc}$ decreases the Higgs doublets become 
tachyonic, giving rise to EW symmetry breaking. 
This is quite similar to the process of EW symmetry breaking in the 
D6-brane models of ref.\cite{imr,cim2}, in which it may be
described as brane recombination of a $b$ brane and a $c$ brane
into a single recombined brane $e$.
Note that, although one-loop positive corrections as given in 
eq.(\ref{massloop}) will in general be present
also for the Higgs fields, one also expects 
large negative contributions from the usual one-loop top-quark
contribution which will again favour EW symmetry breaking
\cite{ir}. 
  
To sum up, the brane content of table \ref{SMatab} give us
an example of how an SM construction can be achieved by
means of intersecting D5-branes. This particular class of
models shares many features already present in the D6-branes
models of \cite{imr}, whereas some important novelties do also appear.
Notice that in this section we have restricted ourselves to only one 
possible quiver configuration of fig. \ref{SMfig}. Some other
inequivalent constructions can also be performed from the rest of
the quivers in that figure, their discussion being postponed to
Appendix II.


\subsection{D5 Left-Right Symmetric Models}

Quite analogously, the LR structure can also be implemented 
in a D5-brane construction. To show this, let us again 
consider a $\ent_3$ orbifold. Since the chirality pattern
is the same for both SM and LR configurations, the possible brane 
distributions will again be those of figure \ref{SMfig}.
Let us consider now the quiver $a_2$).
The brane content with LR spectrum for such quiver 
is shown in table \ref{LRa2tab}.
\begin{table}[htb]
\footnotesize
\renewcommand{\arraystretch}{2}
\begin{center}

\begin{tabular}{|c||c|c|c|}
\hline
$N_i$ & $(n_i^1,m_i^1)$ & $(n_i^2,m_i^2)$ & $\gamma_{w,i}$ \\
\hline\hline
$N_a=3$ & $(n_a^1, \ep\b^1)$ & $(1/\rho, -\oh\ep\tilde\ep)$ & $1_3$ \\
\hline
$N_b=2$ & $ (1/\b^1, 0)$ & $(\tilde\ep, -\frac{3\rho}{2}\ep)$ & $\a 1_2$ \\
\hline
$N_c=2$ & $ (1/\b^1, 0)$ & $(\tilde\ep, -\frac{3\rho}{2}\ep)$ & $\a^2 1_2$ \\
\hline
$N_d=1$ & $(n_d^1, \ep\b^1/\rho )$ & $(1, \frac{3\rho}{2}\ep\tilde\ep)$ 
& $1$ \\
\hline
$N_h$ & $(\ep_h/\b^1,0)$  &  $(2,0)$ & $1_{N_h}$  \\
\hline
\end{tabular}
\caption{\small{D5-branes wrapping numbers and CP phases yielding a LR spectrum
in the $\ent_3$ orbifold of fig.\ref{SMfig}.$a_2$). Solutions are parametrized
by $n_a^1, n_d^1 \in \ent$, $\ep, \tilde\ep = \pm 1$, 
$\beta^1=1-b^{(1)} = 1, 1/2$ and $\rho=1,1/3$.}
\label{LRa2tab}}
\end{center}
\end{table}

Notice that branes $b$ and $c$ belong in fact to the same stack
of four branes, with a non-trivial CP action on it. From the point
of view of gauge fields, however, each one is a separate sector.
Tadpole cancellation conditions are, as usual, almost satisfied when 
imposing this wrapping numbers. The only non-trivial conditions that
remains is
\beq
\frac{3n_a^1}{\rho} \ -\ \frac{2 \tilde\ep}{\b^1} \ +\ n_d^1\ 
+\ 2 N_h \frac{\ep_h}{\beta^1} \ = \ -8 .
\label{tadLRZ3}
\eeq

On the other hand, we must also compute the couplings to RR twisted 
fields, which in this case are
\beq
\begin{array}{rcl}
B_2^{(1)} & \wedge  & \ c_1 (\a -\a^2) {{2\tilde\ep}\over {\b^1} }
(F^{b} - F^{c}) \\
D_2^{(1)} & \wedge  & \ c_1 \left( {{3\rho}\ep \over {\b^1}}\ 
(F^{b} \ +\ F^c) 
-\ 3\ep\tilde\ep (n_a^1 F^a\ - \ \rho  n_d^1 F^d) \right) \
\nonumber \\
E_2^{(1)} & \wedge  &  \ c_1 \frac{2\ep\b^1}{\rho}(3F^a\ +\  F^{d})
\end{array}
\label{bfsLRZ3}
\eeq

This $B \wedge F$ couplings will again give mass to three of
the four $U(1)$ gauge bosons initially present in our spectrum.
If we impose the condition
\beq
n_a^1 \ =\ - 3 \rho n_d^1,
\label{condLRZ3}
\eeq
then the only generator with null coupling to these fields is
$Q_0 = Q_a - 3Q_d$, which corresponds to $U(1)_{B-L}$. After
imposing this condition, tadpoles (\ref{tadLRZ3}) become
\beq
4 n_d^1  \ + \ \frac{1}{\beta^1} \left(\tilde\ep - N_h \ep_h \right) \ = \ 4,
\label{tadlrZ3b}
\eeq
so the extra brane $h$ will be generically necessary in order to 
satisfy tadpoles. 

For completeness, let us give an explicit solution of (\ref{tadlrZ3b}).
Consider the following choice of parameters:
\beqa
n_d^1 = N_h = 1 & \stackrel{(\ref{condLRZ3})}\Longrightarrow 
& n_a^1 = -3\rho \nonumber \\
& \ep_h = \tilde \ep,
\label{choice2}
\eeqa
which now give us a non-trivial $h$ sector with gauge group $U(1)$.
Following the same considerations as in the previous SM construction,
we see that the angles the branes $a$, $d$ and $h$ form with the 
orientifold plane are
\beq
\begin{array}{cc}
\theta_a^1 = \ep\ \left(\pi - tg^{-1}
\left(\frac{\b^1}{3\rho}U^1\right) \right) & 
\theta_a^2 = - \ep\tilde\ep\  tg^{-1}\left(\frac{\rho}{2}U^2\right) \\
\theta_d^1 = \ep\ tg^{-1}\left(\frac{\b^1}{\rho}U^1\right) &
\theta_d^2 = \ep\tilde\ep\  tg^{-1}\left(\frac{3\rho}{2}U^2\right) \\ 
\th_h^1 = \frac{\pi}{2} (1 - \tilde\ep) & \th_h^2 = 0
\label{angulillos2}
\end{array}
\eeq
where again $U^i = R_2^i/R_1^i$, $i=1,2$. Under the choice
$U^1 = \frac{3\rho^2}{2\b^1} U^2$, some of the potential tachyons 
in these sectors will become massless, as for instance those arising from
($aa^*$) intersections. However, just as in the previously discussed 
SM construction some tachyons will remain at ($ad$), ($ad^*$) 
intersections, and some other new tachyons involving the brane $h$.
Again, as in the previous SM case, one-loop contributions 
to the scalar masses may easily overcome the tachyonic contribution.


One can also find an interesting family of left-right symmetric 
models with no open string tachyons already at the tree-level. 
Indeed, it is quite easy to generalize the Left-Right
symmetric spectrum for a $\ent_N$ orbifold with odd $N > 3$.
As an example, let us take the brane content of table 
\ref{wnumbersZN}, which corresponds to a particular case of
fig. \ref{2quivers}.$a$), and that will again give us the spectrum
of table \ref{specLR}.
\begin{table}[h]
\footnotesize
\renewcommand{\arraystretch}{2}
\begin{center}

\begin{tabular}{|c||c|c|c|}
\hline
 $N_i$  &  $(n_i^1,m_i^1)$  &  $(n_i^2,m_i^2)$  & $\gamma_{w,i}$ \\
\hline\hline
$N_a=3$ & $(1/\b^1,0)$ & $(\ep,\oh\tilde\ep)$ & $1_3$ \\
\hline
$N_b=2$ & $(n_b^1,-\tilde\ep\b^1)$ & $ (3, \oh\ep\tilde\ep)$ & $\a 1_2$ \\
\hline
$N_c=2$ & $(n_c^1,\tilde\ep\b^1)$ & $ (3, \oh\ep\tilde\ep)$ & $\a 1_2$ \\
\hline
$N_d=1$ & $(1/\b^1,0)$ & $(-3\ep, \oh\tilde\ep)$ & $1$ \\
\hline
\end{tabular}
\caption{\small{D5-branes wrapping numbers and CP phases yielding a LR spectrum
in a $\ent_N$. Solutions are parametrized
by $n_b^1, n_c^1 \in \ent$, $\ep, \tilde\ep = \pm 1$ and 
$\beta^1=1-b^{(1)} = 1, 1/2$.}
\label{wnumbersZN}}
\end{center}
\end{table}
As in our previous LR example, tadpoles will be cancel by means
of a hidden-brane sector, which in this $\ent_N$ case will
consist of a brane system as shown in table \ref{hiddenZN}.
\begin{table}[htb]
\footnotesize
\renewcommand{\arraystretch}{2}
\begin{center}

\begin{tabular}{|c||c|c|c|}
\hline
$N_i$ & $(n_i^1,m_i^1)$ & $(n_i^2,m_i^2)$ & $\gamma_{w,i}$ \\
\hline\hline
$N_{h1}$ & $(\ep_{h1}/\b^1,0)$ & $(n_h^2,m_h^2)$ & $\a$  \\
\hline
$N_{h2}$ & $(\ep_{h2}/\b^1, 0)$ & $(2,0)$ & $\a^3$  \\
\hline
$\vdots$ & $\vdots$ & $\vdots$  & $\vdots$  \\
\hline
$N_{hs}$ & $(\ep_{hs}/\b^1, 0)$ & $(2,0)$ & $\a^{2s-1}$  \\
\hline
\end{tabular}

\caption{\small {Hidden brane system in a ${\bf Z_N}$ orbifold singularity.}
\label{hiddenZN}}
\end{center}
\end{table}
There one has $\ep_{hi} = \pm 1$ and the value of $s$ is fixed by
tadpole conditions. Consistency conditions in (\ref{tadpoleO5b})
are now easily satisfied. Indeed, second and fourth conditions
are already satisfied with this brane content, while the third
amounts to imposing
\beq
\ep \tilde\ep (n_b^1 + n_c^1) + \ep_{h1} N_{h1} {m_h^2 \over \beta^1} = 0.
\label{tadLRZNa}
\eeq
As mentioned above, the first of these conditions can be expressed as
(\ref{generadortwist}), from where we can read that we must also impose
\beqa
 6 (n_b^1 + n_c^1) + \ep_{h1} N_{h1} {n_h^2 \over \beta^1} = \eta 16 
\label{tadLRZNb} \\
\ep_{hi} N_{hi} {2 \over \beta^1} = \eta 16, \  \ {\rm (i=2,\cdots,r)},
\label{tadLRZNc}
\eeqa
where $r$ and $\eta$ have been defined in (\ref{eta}). Last condition
actually implies $s = r$, $\ep_{hi} = \eta$ and $N_{hi} = 8\b^1$, for 
$i > 1$. Let us also compute the couplings to RR 2-form twisted fields
which will render some of these $U(1)$ gauge bosons massive.
Even if there are in principle 2($N-1$) such fields, most of their
couplings are redundant, so we will still have some massless $U(1)$'s 
in our gauge group. Indeed, these couplings are

\beq
\begin{array}{rcl}
B_2^{(k)} & \wedge & c_k \left( (\alpha^k - \bar\alpha^k)  
\left[ 6(n_b^1 F^b + n_c^1 F^c) + 
\ep_{h1} N_{h1} {n_h^2 \over \beta^1} F^{h1} \right]
+ \sum_{i=2}^r (\alpha^{ik} - \bar\alpha^{ik}) \eta 16 F^{hi} \right) \\
C_2^{(k)} &\wedge & c_k (\alpha^k - \bar\alpha^k) \ep\beta^1 (- F^b + F^c) 
\\
D_2^{(k)} &\wedge & c_k 
\left( {2\tilde\ep \over \beta^1} (3F^a + F^d) + (\alpha^k + \bar\alpha^k) 
\left[\ep\tilde\ep (n_b^1 F^b + n_c^1 F^c) 
+ \ep_{h1} N_{h1} {m_h^2 \over \beta^1} F^{h1} \right] \right)
\\
E_2^{(k)} & \wedge & \ c_k (\alpha^k + \bar\alpha^k) 6\b^1 
\tilde\ep (- F^b + F^c) 
\end{array}
\label{bfslrZN}
\eeq

Imposing tadpole conditions (\ref{tadLRZNa}), (\ref{tadLRZNb})
and (\ref{tadLRZNc})
is easy to see that the only linear combination of abelian groups
that does not couple to any RR field is just 
$U(1)_{B-L} = \frac 13 U(1)_a - U(1)_d$, providing us with
another example of Left-Right symmetric model.
This family of configurations
yielding the same spectrum for arbitrary odd-ordered $\ent_N$
orientifold seems quite interesting, since it gives us
a family of $\ent_N$ models with $N$ arbitrarily large.
In addition they may have an 
open-string tachyonless spectrum. For instance, by the choice of
discrete parameters
\beq
\begin{array}{ll}
n_b^1 = n_c^1 =\eta, & N_{h1} = 4 \b^1, \\ 
\ep_{h1} = \eta, & (n_h^2, m_h^2) = (1, -\oh\ep\tilde\ep),
\end{array}
\label{solZN}
\eeq
conditions (\ref{tadLRZNa}), (\ref{tadLRZNb}) and (\ref{tadLRZNc}) are 
satisfied,
and the compactification radii can also be chosen to avoid
any tachyonic excitation. Indeed, our potential tachyons
will arise only from ($bh1$) and ($ch1$) intersections whose
characteristic angles are
\beq
\pi |\vt_{bh1}^1| = \pi |\vt_{ch1}^1| =
tg^{-1}\left(\b^1 U^1 \right)\ ;\
\pi |\vt_{bh1}^2| = \pi |\vt_{ch1}^2| =
tg^{-1}\left(\frac{U^2}{6}\right) 
+ tg^{-1}\left(\frac{U^2}{2}\right),
\eeq
so by appropriately choosing the complex structure moduli we can
achieve $|\vt_{bh1}^1| = |\vt_{bh1}^2|$ and 
$|\vt_{ch1}^1| = |\vt_{ch1}^2|$, finding
a one-parameter family of tachyonless open-string spectra.

Let us end this subsection by recalling an apparent phenomenological 
shortcoming of the class of left-right symmetric models built here.
Eventually we would like to break the gauge symmetry down to
the Standard Model one and, in order to do that, we need to give a vev to 
a right-handed doublet of scalars with non-vanishing
lepton number. No such scalars are present in the lightest
spectrum of the particular models constructed here.
It would be interesting to find other examples in which
correct gauge symmetry breaking is feasible.

\section{Some physics issues}

\subsection{Low-energy spectrum beyond the SM}

Let us summarize the lightest (open string) spectra
in the class of SM D5-brane constructions:

\begin{itemize}

\item {\it Fermions}
 
The only massless fermions are the ones 
of the SM (plus right-handed neutrinos). In particular,
unlike the case of D6-branes, there are no gauginos 
in the lightest spectrum. 

\item{\it Gauge bosons} 

There are only the ones of the SM
(or its left-right extension). There are in addition three
extra  massive (of order the string scale)
$Z_0$'s, two of them anomalous and the other
being the extra $Z_0$ of left-right symmetric models. 
As discussed in ref.\cite{giiq} for a string scale 
of order a few TeV the presence of these extra $U(1)$'s 
may be amenable to experimental test. In fact already 
present constraints from electroweak
precision data (i.e., $\rho $-parameter) put important
bounds on the mass of these extra gauge bosons.

\item{\it Scalars in the D5-branes  bulk}

There are two copies of scalars in the adjoint representation
of $SU(3)\times SU(2)\times U(1)_a\times U(1)_b\times U(1)_c\times
U(1)_d$, as  
given in eq.(\ref{espectro5aa}). These will include a couple of 
colour octets and $SU(2)_L$ triplets plus eight singlets.
The vevs of the latter parametrize the  locations of 
the four stacks of branes along the two tori
($4\times 2$ parameters) and hence are moduli
at the classical  level. The colour octets and $SU(2)$
triplets get masses at one loop as given in eq.(\ref{massloop}).

\item {\it Scalars at the intersections} 

These are model 
dependent. In the SM example described in some detail 
in section (3.1) there are colour  triplets and sextets 
(from $(aa*)$) and  colour triplets {\it ` leptoquarks'} 
(from $(ad), (ad^*)$) (see table \ref{escalares}). 
 Again their leading contribution to their masses 
should come from eq.(\ref{massloop}). These scalars are
not stable particles, they 
decay into quarks and leptons through Yukawa couplings.
In the SM examples in Appendix II the scalars at the intersections 
are colour singlets.

\item {\it SM Higgs doublets}   

There are sets of Higgs doublets as in table \ref{higgsses}
with a multiplicity which is model dependent. In the example of
section (3.1) the multiplicity is one and hence
we have the same minimal Higgs sector as in the MSSM.

\end{itemize}

The above states constitute the lightest states in the brane 
configuration. At the massive level there will appear 
Kaluza-Klein replicas for the gauge bosons as well as
stringy winding and oscillator states (gonions). 
Compared to the spectra of D6-brane intersection models
\cite{imr,cim1,cim2}  
the present spectrum is quite simpler, since the fermions and
gauge bosons of the SM do not have any kind of SUSY partner.

Note that the structure of the $U(1)$ gauge bosons in D5-brane
models is remarkably similar to that of the D6-brane models of 
ref.\cite{imr,cim1,cim2}. This similarity is dictated by 
the massless chiral fermion spectrum in both classes of 
models which is identical, i.e., the fermions of the SM.
In particular baryon number is a gauged 
symmetry ($U(1)_a$) which remains as a global symmetry in
perturbation theory once the corresponding $U(1)$'s become massive.
This naturally guarantees proton stability.

Concerning the closed string sector, the $\ent_N$ 
projection kills all fermionic partners of the 
untwisted sector. We will have the graviton plus
a number of untwisted moduli field as well as 
untwisted RR-fields. The twisted closed string sector
is relevant to anomaly cancellation.

\subsection{Lowering the string scale}

\begin{figure}
\centering
\epsfxsize=4in
\hspace*{0in}\vspace*{.2in}
\epsffile{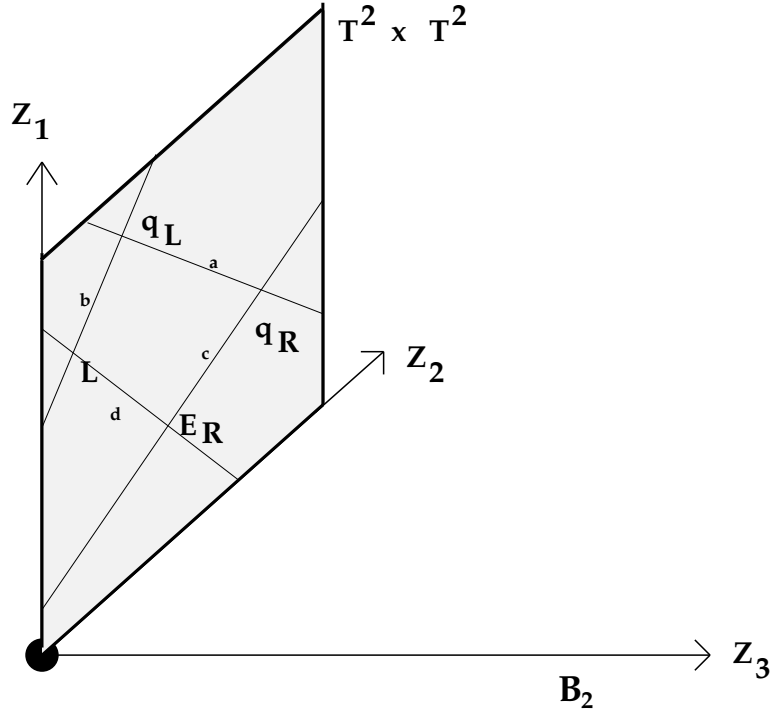}
\caption{\small Intersecting D5-world set up. 
The $Z_i$, $i=1,2,3$ represent complex compact dimensions.
The  D5-branes $a,b,c,d$ (corresponding to the gauge group
$U(3)\times U(2)\times U(1)\times U(1)$) wrap cycles
on ${\bf T^2\times T^2}$. At the intersections lie 
quarks and leptons. This system is transverse to a
2-dimensional compact space  ${\bf B_2}$ (e.g., ${\bf T^2/Z_N}$)
whose volume may be quite large so as to explain 
$M_p>>M_s$. This would be a D-brane realization of the 
scenario in ref.\cite{aadd}.}
\label{braneworld}  
\end{figure}

The D5-brane models here constructed are non-supersymmetric.
In order not to have the standard hierarchy problem for
Higgs scalars the most obvious possibility is to have 
the fundamental string scale not much above 
the weak scale. Thus we  should have $M_s\propto 1-10$ TeV.

Interestingly enough,
in the intersecting D5-brane models here studied
one can  have a low string scale $M_s\propto 1-10$ TeV
while maintaining the experimentally measured four-dimensional
Planck mass $M_p= 1.18 \times 10^{19}$ GeV by
some dimensions getting very large \cite{aadd}.
Indeed, in the present examples the compact space
has the form ${\bf T^4\times B_2}$, and the 
D5-branes sit at a ${\bf C/Z_N}$ singularity in
${\bf B_2}$  and wrap two-cycles
on ${\bf T^4}$. Let us denote by $V_4$ the volume of ${\bf T^4}$ and by
$V_2$ that of the manifold ${\bf B_2}$. Then the Planck scale
is given by 
\beq
M_p \ =\  {2\over \lambda }  M_s^4 \sqrt{V_4V_2} 
\eeq
In order to avoid too light KK/Winding modes in the 
worldvolume of the D5-branes let us assume $V_4\propto 1/M_s^4$.
Then one has 
\beq
V_2 \ =\  {  { M_p^2 \lambda ^2 } \over {4 M_s^4} }
\eeq
and one can accommodate a low string scale $M_s\propto 1$ TeV
by having the volume $V_2$  of the 2-dimensional manifold ${\bf B_2}$
large enough (i.e., of order $(mm.)^2$). For a pictorial view of
this explicit D-brane realization of the proposal in \cite{aadd}
see fig.\ref{braneworld}.

\section{Final comments}

In this paper we have presented D5-brane configurations 
wrapping cycles on ${\bf T^2\times T^2\times}$ $({\bf C/\ent_N})$ yielding
the massless fermionic spectrum of the three-generation SM.
This is a generalization of the work in ref.\cite{imr}
in which it was obtained the SM spectrum from D6-branes 
wrapping cycles on ${\bf T^2\times T^2\times T^2}$. 
We have also presented for completeness the case of D4-branes 
wrapping cycles on ${\bf T^2\times (T^4/Z_N)}$, which turns out
to be less flexible from the model-building point of view.

One of our main motivations to consider the case of D5-branes 
is the fact that in this case there are  2 dimensions which are 
transverse to  the SM D5-brane configuration. By making those
two dimensions large enough one can have a low string scale 
$M_s$ of order 1-10 TeV and still have a large $M_{Planck}$
in agreement with observations. From this point of view 
these are the first explicit D-brane string constructions in which
one has just the fermionic spectrum of the SM at low energies
and the mechanism for lowering the string scale in
\cite{aadd} simultaneously at work.

There are a number of questions both theoretical and 
phenomenological which we have not addressed in this paper 
and should be the subject of further research. 
These D5-brane constructions are non-supersymmetric and
it remains to be seen if such configurations can 
be rendered stable. One source of instability may be 
the presence of closed string tachyons in the twisted 
spectrum. An analogous class  of tachyons have been studied recently 
in ref.\cite{tachyons} in the non-compact orbifold  
case. That analysis cannot be directly traslated to 
the compact orientifold case considered here in which e.g. the
tachyons are real rather than complex fields.
It remains to be seen
whether in the compact {\em orientifold case}  a stabilization
of the
closed string tachyons may be feasible.
 This is also relevant to the question
whether one can obtain a stable minimum in which the two
dimensions transverse to the SM brane configuration are very
large compared to the rest, thus providing for a dynamical
explanation of the smallness of the string scale compared to
the Planck mass.

The only fermions in the light spectrum are those of
the non-SUSY SM. The fermions do not have any 
SUSY-partners, no squarks or sleptons appear.
There are however some scalars in the lightest spectrum.
 There are some with the quantum numbers of
electroweak Higgs fields which may become tachyonic and trigger 
electroweak symmetry breaking if certain branes are sufficiently close.
On the other hand there are further scalars which may be tachyonic at the 
tree level.  In the simple SM example in the text those 
are coloured particles and we argue
that their full mass$^2$ including one-loop effects will in general be
positive. Those coloured (triplets and sextets ) should then be
relatively light with masses close to the electroweak scale. They are
unstable and decay into ordinary quarks and leptons. In addition there are
three extra $Z_0$'s beyond the ordinary one with masses of order the string scale
(i.e., 1-10 TeV in low string models). These may lead already to
observable effects as recently argued in ref.\cite{giiq}.
The fact that baryon number is gauged will guarantee that in these 
constructions the proton is perturbatively stable.
We leave a more systematic study of the phenomenological aspects 
of this class of brane models for future work.

\vspace{1cm}

\centerline{\bf Acknowledgements}
We are grateful to G. Aldaz\'abal, A. Font, C. Kokorelis,  
R. Rabad\'an and  A.~Uranga for useful discussions.
The research of D.C.  and F.M. was  supported by
 the Ministerio de Educaci\'on, Cultura y Deporte (Spain) through FPU grants.
This work is partially supported by CICYT (Spain) and the
European Commission (RTN contract HPRN-CT-2000-00148).

\newpage

\section{Appendix I: D4-branes wrapping on 
${\bf T^2\times C^2/\ent_N}$ orientifolds }

For the sake of completeness, in this appendix we present
the general construction involving intersecting D4-branes
in an orientifold singularity. This general class of models
is both of theoretical and phenomenological interest
since they also provide a natural setup for considering
chiral compactifications with low string scale scenarios.
To be concrete, we will consider the compactification
\beq
{\rm Type \ IIA \ on \ } M_4 \ti 
\frac{\ T^{2} \ti \C^2/\ent_N}{\{1 + \OR\}},
\label{singuoriD4}
\eeq
where $\R$ now stands for $\R_{(5)}\R_{(6)}\R_{(7)}\R_{(8)}\R_{(9)}$.
In terms of its action on complex coordinates this involution 
is given by
\beqa
{\cal R}: & Z_1 \longmapsto \bar Z_1, \\
& Z_i \longmapsto - Z_i, & i = 2, 3.
\label{R4}
\eeqa
This theory will contain a O$4$-plane wrapping a 1-cycle 
in ${\bf T^2}$ (the one invariant under $\R_{(5)}$, and in order to 
cancel its negative RR charge we will have to include an
open string sectors involving D$4$-branes wrapping 1-cycles 
$[\Pi] = [(n,m)]$ \footnote{Notice that in this 
particular class of compactifications every cycle is 
factorizable.} of this same ${\bf T^2}$, while sitting at 
the origin of $\C^2/\ent_N$.

The geometric action of the orbifold group $\ent_N$ 
can be described by a twist vector 
$v_\om = {1 \over N} (0,b_1,b_2,0)$,
$b_1 = b_2$ mod $2$ for the variety to admit spinors.
This twist will preserve some bulk supersymmetry
whenever $b_1 = \pm b_2$ mod $N$.  
Just as in the case of D5-branes, the orbifold 
action on the Chan-Paton degrees of freedom can 
be described by a matrix of the form (\ref{Chan}),
and the orientifold action can also be implemented
by adding a mirror sector $a^*$ for every D4-brane
$a$ in the configuration. If we again consider effective
wrapping number for describing our 1-cycles, mirror
branes will be related in an analogous way that
the one described in (\ref{brana_a}) and (\ref{brana_a*})
for the case of D$5$-branes.

Let us now describe the low energy spectrum of the theory

\begin{itemize}

\item{Closed string sector}

The twisted closed string sector will consist, 
in the supersymmetric case,
of a $D = 4$ $\cn = 4$ $U^{(\frac{N-1}{2})}$
gauge multiplet for odd $N$, the gauge group
being $U^{(\frac{N}{2})}$ if $N$ is even.
When dealing with the non-supersymmetric $|b_1| \neq |b_2|$ 
case, however, the twisted closed string spectrum
will be much similar to the case of D$5$-branes,
a closed string tachyon appearing for each twisted sector.

\item{$D4_aD4_a$ sector}

This sector gets mapped to $D4_{a^*}D4_{a^*}$, which 
usually is a different sector of the theory. The computation
of its massless spectrum will be the same as in the orbifold case,
already computed in \cite{afiru}. However, we present 
its computation for completeness.
The massless GSO projected states in both R and NS sectors are
{\small \beq
\begin{array}{cccc}
{\rm \bf NS \ Sector} \quad & \quad {\bf \ent_N \ phase}  \quad 
& \quad {\rm \bf R \ Sector} \quad & \quad {\bf \ent_N \ phase} \\
(\pm1,0,0,0) & 1 & \pm\oh(-,+,+,+) & e^{\pm\pi i \frac{b_1 + b_2}{N}}\\
(0,\pm1,0,0) & e^{\pm2\pi i \frac{b_1}{N}} 
& \pm\oh(+,-,+,+) & e^{\mp\pi i \frac{b_1 - b_2}{N}}\\
(0,0,\pm1,0) & e^{\pm2\pi i \frac{b_2}{N}} 
& \pm\oh(+,+,-,+) & e^{\pm\pi i \frac{b_1 - b_2}{N}}\\
(0,0,0,\pm1) & 1 & \pm\oh(+,+,+,-) & e^{\pm\pi i \frac{b_1 + b_2}{N}}
\label{sector4aa}
\end{array}
\eeq}
where the behaviour of such states under the $\ent_N$ action 
has been indicated. Keeping states invariant under combined 
geometrical and Chan-Paton action we are left with the 
following spectrum
{\small\beq
\begin{array}{rl}
{\rm\bf Gauge\; Bosons} & \quad \prod_a \prod_{i=1}^N U(N_a^i) \\
{\rm\bf Complex\; Escalars} & \quad \sum_a \sum_{i=1}^N \;
[\, (N_a^i,{\ov N}_a^{i+b_1}) + (N_a^i,{\ov N}_a^{i+b_2}) \, 
+ {\bf Adj}_a^i ] \\
{\rm\bf Left\; Fermions} & \quad \sum_a \sum_{i=1}^N \; [\;
(N_a^i,{\ov N}_a^{i-(b_1-b_2)/2}) + (N_a^i,{\ov N}_a^{i+(b_1-b_2)/2}) ]
\nonumber \\
{\rm\bf Right\; Fermions} & \quad \sum_a \sum_{i=1}^N \; [\;
(N_a^i,{\ov N}_a^{i+(b_1+b_2)/2}) + (N_a^i,{\ov N}_a^{i-(b_1+b_2)/2})]
\label{espectro4aa}
\end{array}
\eeq}
which is generically non-supersymmetric and always non-chiral.
The supersymmetric twist give us the $\cn = 2$ theory
{\small\beq
\begin{array}{rl}
{\rm\bf Vector\; Multiplet} &\quad \prod_a \prod_{i=1}^N U(N_a^i)
\nonumber\\
{\rm\bf Hypermultiplet} & \quad \sum_a \sum_{i=1}^N (N_a^i,{\ov
N}_a^{i+1})
\label{multipletes4aa}
\end{array}
\eeq}

\item{$D4_aD4_b$, $D4_aD4_{b^*}$ and $D4_aD4_{a^*}$ sectors}

These three sectors will contain the chiral spectrum of the theory. 
Let us analyze the $D4_aD4_b$ spectrum, whose associated twisted
vector is given by $v_\vt = (0,\vt_{ab},0,0)$. Being mapped into
$D4_{b^*}D4_{a^*}$ under the action of $\OR$, we only have to consider 
the orbifold action. The massless states are
{\small\beq
\begin{array}{cccc}
{\rm\bf Sector} & {\rm\bf State} & {\bf \ent_N \ phase}  
& {\rm\bf \a^\prime Mass^2} \\
{\rm NS} & (-1+\vartheta,0,0,0) & 1 & -\oh |\vt_{ab}|\\
{\rm R}  & (-\oh+\vartheta,+\oh,-\oh,-\oh) &
e^{\pi i\frac{(b_1-b_2)}{N}} & 0 \\
         & (-\oh+\vartheta,-\oh,+\oh,-\oh) &
e^{-\pi i\frac{(b_1-b_2)}{N}} & 0 \\
         & (-\oh+\vartheta,-\oh,-\oh,+\oh) & 
e^{-\pi i\frac{(b_1+b_2)}{N}} & 0 \\
         & (-\oh+\vartheta,+\oh,+\oh,+\oh) & 
e^{\pi i\frac{(b_1+b_2)}{N}} & 0
\end{array}
\label{sector4ab}
\eeq}
where we have supposed $0 < \vt_{ab} < 1$. This spectrum is 
explicitely non-supersymmetric, even for a supersymmetric twist.
Keeping the invariant states we are left with the spectrum
{\small\beq
\begin{array}{rl}
{\rm\bf Tachyons} & \sum_{a<b} \sum_{i=1}^N \; I_{ab}\times
(N_a^i,{\ov N}_b^i) \\
{\rm\bf Left\; Fermions} & \sum_{a<b} \sum_{i=1}^N \;  I_{ab}\times\;
[\; (N_a^i,{\ov  N}_b^{i-(b_1+b_2)/2}) + (N_a^i,{\ov N}_b^{i+(b_1+b_2)/2})
\; ] \\
{\rm\bf Right\; Fermions} & \sum_{a<b} \sum_{i=1}^N \;I_{ab}\times
[\; (N_a^i,{\ov  N}_b^{i-(b_1-b_2)/2}) + (N_a^i,{\ov  N}_b^{i+(b_1-b_2)/2})
\; ]
\end{array}
\label{espectro4ab}
\eeq}
which is generically supersymmetric. Notice that the intersection
number is now given by 
$I_{ab} \equiv [\Pi_a]\cdot[\Pi_b] = n_am_b - m_an_b$. Similarly,
we can compute the other two chiral sectors of the theory,
the complete spectrum being
\footnote{In case $n_a = 0$, there is just one tachyon coming from 
the $aa^*$ sector, transforming in the 
antisymmetric representation $({\bf A}_a^0)$ 
of the $U(N_a^0)$ gauge group. This is just a T-dual orbifolded version of
the non-BPS systems constructed in \cite{antisym}.}
{\small\beq
\begin{array}{l}
{\rm\bf Tachyons} \\ \sum_{a<b} \sum_{i=1}^N \; 
[\; I_{ab}(N_a^i,{\ov N}_b^i) + 
I_{ab^*}(N_a^i,N_b^{-i})\;]\\
\sum_a [\;2 |m_a|(|n_a| + 1) ({\bf A}_a^0) 
+ 2 |m_a|(|n_a| - 1) ({\bf S}_a^0)\;]
\vspace{3mm}\\ 
{\rm\bf Left\; Fermions}  \\ \sum_{a<b} \sum_{i=1}^N \; 
I_{ab} [\; (N_a^i,{\ov  N}_b^{i-\oh{(b_1+b_2)}}) 
+ (N_a^i,{\ov N}_b^{i+\oh{(b_1+b_2)}}) \;]\\
\sum_{a<b} \sum_{i=1}^N \; 
I_{ab^*} [\; (N_a^i,N_b^{-i+\oh{(b_1+b_2)}}) 
+ (N_a^i,N_b^{-i-\oh{(b_1+b_2)}}) \;] \\
\sum_a \sum_{j,i=1}^N \; (\d_{j,-i+\oh(b_1+b_2)}+\d_{j,-i-\oh(b_1+b_2)})
[\; - m_a (n_a + 1) ({\bf A}_a^j) - m_a (n_a - 1) ({\bf S}_a^j)\;]
\vspace{3mm}\\
{\rm\bf Right\; Fermions} \\ \sum_{a<b} \sum_{i=1}^N \; 
I_{ab} [\; (N_a^i,{\ov  N}_b^{i-\oh{(b_1-b_2)}}) 
+ (N_a^i,{\ov N}_b^{i+\oh{(b_1-b_2)}}) \;]\\
\sum_{a<b} \sum_{i=1}^N \; 
I_{ab^*} [\; (N_a^i,N_b^{-i+\oh{(b_1-b_2)}}) 
+ (N_a^i,N_b^{-i-\oh{(b_1-b_2)}}) \;]\\
\sum_a \sum_{j,i=1}^N \; (\d_{j,-i+\oh{(b_1-b_2)}}+\d_{j,-i-\oh{(b_1-b_2)}})
[\; - m_a (n_a + 1) ({\bf A}_a^j) - m_a (n_a - 1) ({\bf S}_a^j)\;]
\end{array}
\label{espectro4ab*}
\eeq}

\end{itemize}

The construction of these configurations are, as usual, constrained by
tadpole cancellation conditions, which in this case read
\beqa
& & c_k^2 \ \sum_a n_a \ \left({\rm Tr} \gamma_{k,a} 
+ {\rm Tr} \gamma_{k,a^*} \right) = 8  
\prod_{r = 1}^2 {\rm sin } \left(\frac{\pi k b_r}{2N}\right),
\label{tadpoleO4n}\\
& & c_k^2 \ \sum_a m_a \left({\rm Tr} \gamma_{k,a} 
- {\rm Tr} \gamma_{k,a^*} \right) = 0,
\label{tadpoleO4m}
\eeqa
where now $c_k^2 = \prod_{r = 1}^2 {\rm sin }(\pi k b_r/N)$,
and we are using effective wrapping numbers.
These two conditions can be expressed more elegantly as
\beq
c_k^2 \ \sum_a \left([\Pi_a] \ {\rm Tr} \gamma_{k,a} 
+ [\Pi_{a^*}] \ {\rm Tr} \gamma_{k,a^*} \right)
= [\Pi_{O4}] \ 8 \b 
\prod_{r = 1}^2 {\rm sin} \left(\frac{\pi k b_r}{2N}\right),
\label{tadpoleO4}
\eeq
where $\b = 1 - b$ discriminates between rectangular and tilted
tori. In the same manner as scketched for the case of D$5$-branes in
section 2, tadpole conditions will directly imply cancellation
of cubic chiral anomalies, whose expression is now given by
\beqa
\ca_{SU(N_a^j)^3} & = & \sum_{b,k} N_b^k \left(I_{ab} \ \d(j,k) 
+ I_{ab^*} \ \d(j,-k)\right) + 8\b \ I_{a,O4}\ \d(j,-j) \\
\d(j,k) & \equiv & \d_{j,k+\frac{b_1+b_2}{2}} 
+ \d_{j,k-\frac{b_1+b_2}{2}} -
\d_{j,k+\frac{b_1-b_2}{2}} - \d_{j,k-\frac{b_1-b_2}{2}}.
\label{quiralO4}
\eeqa

On the other hand, the mixed anomalies analysis mimicks
the one performed in Section 2 for D$5$-branes. In fact,
expressions (\ref{mixtaO5}) and (\ref{mixtaO5b}) are 
also valid for this case if we just substitute 
$16 m_a^1 m_a^2$ by $8 m_a$ and consider the definitions of 
$\d(i,j)$ and $c_k$ used in this appendix. The only 
difference comes from the details of the GS mechanism
which now only involves $(N-1)$ RR twisted fields.
For completeness, we present the four-dimensional couplings
that give rise to such mechanism
\beqa & &
\begin{array}{c}
c_k N_a\, n_a \int_{M_4} {\rm Tr} \left(\g_{k,a}-\g_{k,a^*}\right)\lam_i
\ C_2^{(k)}\wedge {\rm Tr} F_{a,i}, \\
c_k N_a\, m_a \int_{M_4} {\rm Tr} \left(\g_{k,a}+\g_{k,a^*}\right)\lam_i \
B_2^{(k)}\wedge {\rm Tr} F_{a,i},
\end{array} 
\label{acoplosdual2O4}
\\ & &
\begin{array}{c}
c_k N_b m_b \int_{M_4} \left(\g_{k,b}^{-1}-\g_{k,b^*}^{-1}\right)\lam_j^2 \
C_0^{(k)} \wedge {\rm Tr} \left(F_{b,j}\wedge F_{b,j}\right), \\
c_k N_b n_b \int_{M_4} \left(\g_{k,b}^{-1}+\g_{k,b^*}^{-1}\right)\lam_j^2 \
B_0^{(k)} \wedge {\rm Tr} \left(F_{b,j}\wedge F_{b,j}\right).
\end{array}
\label{acoplosdualO4}
\eeqa
Of special interest are the couplings (\ref{acoplosdual2O4}),
which encode the massive $U(1)$'s of the theory.

Let us also scketch some model-building features regarding D4-branes
orientifold models. For simplicity, we will constrain ourselves
to the supersymetric case $b_1 = - b_2 = 1$.
Since the $\ent_3$ orbifold case has already been
considered in \cite{honecker}, we will focus on odd $N > 3$ orientifolds.
In the same way as performed for D5-branes tadpoles, we can
express the tadpole condition (\ref{tadpoleO4n}) as
\beq
\sum_a{n_a \left({\rm Tr} \gamma_{\om,a} 
+ {\rm Tr} \gamma_{\om,a^*}\right)} =
{8 \over \left({\alpha^{N+1 \over 4}  + \bar\alpha^{N+1 \over 4}}\right)^2},
\label{generadortwist2}
\eeq
where $\alpha = e^{2\pi i/N}$. Again
we can reexpress (\ref{generadortwist2}) as a sum of orbifold
phases by using
\beq
{1 \over {\alpha^{N+1 \over 4}  + \bar\alpha^{N+1 \over 4}}} =
\iota \left(1 + \sum_{l=1}^r{(\alpha^{r}+ \bar\alpha^{r})}\right),
\label{decomp2}
\eeq
\beq
\iota = \left\{\begin{array}{l}
+1 \ {\rm if} \ N = 4r+1 \\
-1 \ {\rm if} \ N = 4r+3
\end{array}\right.
\label{iota}
\eeq

Let us, for instance, consider the $\ent_5$ orientifold 
model whose brane content is shown in table \ref{D4model}.
\begin{table}[htb]
\footnotesize
\renewcommand{\arraystretch}{2}
\begin{center}

\begin{tabular}{|c||c|c|}
\hline
$N_i$ & $(n_i,m_i)$ & $\gamma_{w,i}$ \\
\hline\hline
$N_a=3$ & $(2, 0) $ & $\a 1_3$ \\
\hline
$N_b=2$ & $ (1, -\frac 32)$ & $\a^2 1_2$ \\
\hline
$N_c=2$ & $ (-1, \frac 32)$ & $\a^2 1_2$ \\
\hline
$N_d=1$ & $(2, 0) $ & $\a 1_3$ \\
\hline
$N_h = 4$ & $(2, 0)$ & $1_{4}$  \\
\hline
\end{tabular}
\caption{\small{Example of a D4-branes LR model in a $\ent_5$ orbifold.} 
\label{D4model}}
\end{center}
\end{table}
As usual, the brane content of this model consist of four D$4$-branes
$a, b, c, d$, again identified with those of table \ref{SMbranes},
plus some hidden brane $h$. The gauge group is
$SU(3) \ti SU(2) \ti SU(2) \ti U(1)^4 \ti [U(4)_h]$, 
which is the LR gauge group extended by three abelian groups and
one hidden $U(4)_h$. The chiral matter content of such model is 
given in table \ref{D4fermions}
\begin{table}[htb] \footnotesize
\renewcommand{\arraystretch}{1.25}
\begin{center}
\begin{tabular}{|c|c|c|c|c|c|c|c|}
\hline Intersection &
 Matter fields  &   &  $Q_a$  & $Q_b $ & $Q_c $ & $Q_d$  & $B-L$ \\
\hline
\hline ($ab$) & $Q_L$ & $3(3,2,1)$ & 1 & -1 & 0 & 0 & 1/3 \\
\hline ($ac$) & $Q_R$ & $3({\bar 3},1,2)$ & -1  & 0 & 1 & 0 & -1/3 \\
\hline ($bd$) & $L_L$ & $3(1,2,1)$ &  0  & -1 & 0 & 1 & -1 \\
\hline ($cd$) & $L_R$ & $3(1,1,2)$ &  0 & 0 & 1 & -1 & 1 \\
\hline ($bc^*$) & $H$ & $3(1,2,2)$ &  0  & 1 & 1 & 0 & 0 \\
\hline ($bb^*$) & $A_i$  & $3(1,1,1)$ & 0 & -2 & 0 & 0 & 0  \\
\hline ($cc^*$) & $S_i$  & $3(1,1,3)$ & 0 & 0 & -2 & 0 & 0  \\
\hline \end{tabular}
\end{center} \caption{\small Extended Left-Right symmetric chiral 
spectrum arising from the $\ent_5$ D4-branes model of table 
\ref{D4model}. The $U(1)_{B-L}$ generator is defined as 
$Q_{B-L} = \frac 13 Q_a - Q_d$.}
\label{D4fermions}
\end{table}

Notice that this particular example does not 
follow the general philosophy described in Section 3, where
every chiral fermion arised from a bifundamental representation
and the matter content was thus described by table \ref{specLR}.
Instead, we now find some extra chiral fermions that can be 
identified with three Higgssino-like particles, whereas some 
exotic matter transforming as singlets  ($A_i$) and symmetrics 
of $SU(2)_R$  ($S_i$) do also appear. 
The only light bosonic sector arises from branes $b$ and $c$
giving us a Higgs-like particle that can become tachyonic
if we approach both branes close enough.
No extra chiral matter nor scalars arise from the hidden 
sector of the theory.

It is easy to see that this brane content satisfies both twisted
tadpole conditions (\ref{tadpoleO4n}) and (\ref{tadpoleO4m}).
Interestingly enough, it also satisfies untwisted tadpoles
conditions, so when embedding such model in a compact 
four-dimensional manifold $B$ no extra brane content would be 
needed.

Finally, by computing the couplings (\ref{acoplosdual2O4}) that
mediate the GS mechanism, we can check that two of the abelian
gauge groups are in fact massive, the only massless linear combinations
being $U(1)_{B-L} = \frac 13 U(1)_a - U(1)_d$ and
$U(1)_b + U(1)_c$, just as in our $\ent_N$ D$5$-branes constructions
of Section 3.2.

Note that the $\ent_5$ twist in this model preserves a $\cn = 2$ 
supersymmetry of the gravitational bulk. Due to this fact there
are no closed string twisted tachyons.

\newpage

\section{Appendix II: Other D5-brane configurations yielding SM spectra }

Althought in Section 3 we have focussed on a very particular class
of D5-branes configurations in a $\ent_3$ orbifold, there are other
possibilities when constructing models giving rise to just the
SM fermionic spectrum. Indeed, the brane content of table
\ref{SMatab} corresponds to the brane distribution of 
fig.\ref{SMfig}.$a_1$), while in principle any of these four 
figures is valid. For completeness, in this appendix we consider the
other three possibilities. 

After imposing the analogous constraints to the rest of the $\ent_3$
quivers of figure \ref{SMfig}, we find that the distribution 
$a_2$) give us a totally equivalent class of models to the one already 
presented, whereas $b_1$) and $b_2$) give us two new different 
families of configurations.
Let us first consider  the $\ent_3$ quiver in fig. \ref{SMfig}.$b_1$). 
The wrapping numbers giving the same SM spectrum of table \ref{specSM} 
are shown in table \ref{SMbtab}.

\begin{table}[htb] 
\footnotesize
\renewcommand{\arraystretch}{1.7}
\begin{center}

\begin{tabular}{|c||c|c|c|}
\hline
 $N_i$    &  $(n_i^1,m_i^1)$  &  $(n_i^2,m_i^2)$   & $\gamma_{\om,i}$ \\
\hline\hline 
$N_a = 3$ & $(1/\b^1, 0)$  &  $(\ep, -\oh\tilde \ep)$ & 
$\a 1_3$  \\
\hline 
$N_b = 2$ & $(n_b^1, \tilde\ep\b^1)$ &
$(1, -\frac 32 \ep\tilde\ep)$ &  $1_2$   \\
\hline 
$N_c = 1$ & $(n_c^1, 3\ep\b^1)$ & $(0, 1)$ & $1$ \\
\hline 
$N_d=1$ & $(1/\b^1, 0)$ & $(\ep, \frac 32 \tilde\ep) $ & $\a$ \\
\hline
$N_h$ & $(\ep_h/\b^1, 0)$ & $(2,0)$ & $1_{N_h}$  \\
\hline
\end{tabular}

\caption{\small{D5-branes wrapping numbers and CP phases giving rise
to a SM spectrum in the $\ent_3$ quiver of fig. \ref{SMfig}.$b_1$). 
The solution is now parametrized by $n_b^1, n_c^1 \in \ent$, 
$\ep, \tilde\ep = \pm 1$, and $\b^1 = 1 - b^{(1)} = 1, 1/2$.}
\label{SMbtab}}
\end{center}
\end{table}

Just as before, tadpoles are almost automatically satisfied, and the 
only condition to be imposed is
\beq
n_b^1 = -4 + \frac{1}{\b^1} (\ep - N_h\ep_h).
\label{tadSMZ3b}
\eeq

The $U(1)$ structure is quite similar as well, again with three 
non-trivial couplings to RR twisted fields, now given by
\beq
\begin{array}{rcl}
B_2^{(1)} & \wedge  & \ c_1\ (\a -\a^2)\ \frac{\ep}{\b^1}
\ (3F^a\ +\ F^{d})\\
D_2^{(1)} & \wedge  & \ c_1 
\left( \frac{3 \tilde\ep}{2\b^1} (F^a\ -\ F^d)\ -\ 6 n_b^ 1 \ep\tilde\ep F^b\ 
+\ 2 n_c^1 F^c \right) \\
E_2^{(1)} & \wedge  & \ c_1\ 4 \tilde\ep \b^1 F^b
\end{array}
\label{bfsSMZ3b}
\eeq
The massless $U(1)$ will again be model-dependent
\beq
Q_0 = Q_a - 3 Q_d - \frac{3\tilde\ep}{n_c^1\b^1} Q_c,
\label{masslessSMZ3b}
\eeq
and getting the hypercharge as the unique massless $U(1)$ amounts
to requiring
\beq
n_c^1 = \frac {\tilde\ep}{\b^1}\ \Rightarrow\ \b^1 = 1,  
\label{condSMZ3b}
\eeq
since $n_c^1$ has to be an integer.
A simple solution is 
 $N_h = 3$, $\ep = -\ep_h = 1$. 
This implies setting 
 $n_h^1 = 0$, and then we have a single Higgs system as in table
\ref{higgsses}.
 
Considering now the quiver in fig. \ref{SMfig}.$b_2$)
give us another family of configurations. 
Looking for the same spectrum than in table
\ref{specSM}, we find the following wrapping numbers:
\begin{table}[htb] 
\footnotesize
\renewcommand{\arraystretch}{1.7}
\begin{center}

\begin{tabular}{|c||c|c|c|}
\hline
 $N_i$    &  $(n_i^1,m_i^1)$  &  $(n_i^2,m_i^2)$   & $\gamma_{\om,i}$ \\
\hline\hline 
$N_a = 3$ & $(1/\b^1, 0)$  &  $(\ep, -\oh\tilde \ep)$ & 
$\a 1_3$  \\
\hline 
$N_b = 2$ & $(n_b^1, \tilde\ep\b^1)$ &
$(1, -\frac 32 \ep\tilde\ep)$ &  $1_2$   \\
\hline 
$N_c = 1$ & $(n_c^1, 3\ep\b^1)$ & $(0, 1)$ & $1$ \\
\hline 
$N_d=1$ & $(1/\b^1, 0)$ & $(-\ep, -\frac 32 \tilde\ep) $ & $\a^2$ \\
\hline
$N_h$ & $(\ep_h/\b^1, 0)$ & $(2,0)$ & $1_{N_h}$  \\
\hline
\end{tabular}

\caption{\small{D5-branes wrapping numbers and CP phases giving rise
to a SM spectrum in the $\ent_3$ quiver of fig. \ref{SMfig}.$b_2$). 
The solution is now parametrized by $n_b^1, n_c^1 \in \ent$, 
$\ep, \tilde\ep = \pm 1$, and $\b^1 = 1 - b^{(1)} = 1, 1/2$.}
\label{SMb2tab}}
\end{center}
\end{table}

Tadpoles read:
\beq
2 n_b^1 = -8 + \frac{1}{\b^1} (\ep - 2 N_h\ep_h)\ \Rightarrow\ \b^1 = \oh.
\label{tadSMZ3a2}
\eeq

The $U(1)$ couplings are:
\beq
\begin{array}{rcl}
B_2^{(1)} & \wedge  & \ c_1\ (\a -\a^2)\ \frac{\ep}{\b^1}
\ (3F^a\ +\ F^{d})\\
D_2^{(1)} & \wedge  & \ c_1 
\left( \frac{3 \tilde\ep}{2\b^1} (F^a\ {\bf +}\ F^d)\ 
-\ 6 n_b^ 1 \ep\tilde\ep F^b\ 
+\ 2 n_c^1 F^c \right) \\
E_2^{(1)} & \wedge  & \ c_1\ 4 \tilde\ep \b^1 F^b
\end{array}
\label{bfsSMZ3a2}
\eeq

The massless $U(1)$ will now be
\beq
Q_0 = Q_a - 3 Q_d + \frac{3\tilde\ep}{2 n_c^1\b^1} Q_c,
\label{masslessSMZ3a2}
\eeq
and getting the hypercharge as the unique massless $U(1)$ amounts
to requiring
\beq
n_c^1 = - \frac {\tilde\ep}{2\b^1} = - \tilde\ep.
\label{condSMZ3a2}
\eeq
Unlike the SM D5-brane constructions in the main text, 
the lightest scalars and/or tachyons are now coulour singlets.

\end{document}